\documentclass[iop]{emulateapj}

\newcommand{\gps}{\ensuremath{g_{\rm P1}}}
\newcommand{\rps}{\ensuremath{r_{\rm P1}}}
\newcommand{\ips}{\ensuremath{i_{\rm P1}}}
\newcommand{\zps}{\ensuremath{z_{\rm P1}}}
\newcommand{\yps}{\ensuremath{y_{\rm P1}}}

\def\ra#1#2#3{#1$^{\rm h}$#2$^{\rm m}$#3$^{\rm s}$}
\def\dec#1#2#3{$#1^\circ#2'#3''$}

\usepackage{epsfig,graphicx}
\usepackage{mathtools}
\usepackage{natbib}
\usepackage{comment}
\usepackage{url}
\usepackage[usenames,dvipsnames]{color}
\usepackage[normalem]{ulem}

\begin{document}

\title{Hydrogen-Poor Superluminous Supernovae from the Pan-STARRS1 Medium Deep Survey}
\submitted{ApJ in press}
\email{ragnhild.lunnan@astro.su.se}

\def\okc{1}
\def\cfa{2}
\def\ou{3}
\def\jhu{4}
\def\stsci{5}
\def\kipac{6}
\def\mit{7}
\def\car{8}
\def\ucsc{9}
\def\nwu{10}
\def\nrao{11}
\def\ucb{12}
\def\noao{13}
\def\gem{14}
\def\kicp{15}
\def\qub{16}
\def\ifa{17}
\def\dur{18}

\author{R.~Lunnan\altaffilmark{\okc,\cfa},
 R.~Chornock\altaffilmark{\ou},
 E.~Berger\altaffilmark{\cfa},
 D.~O.~Jones\altaffilmark{\jhu},
 A.~Rest\altaffilmark{\stsci},
 I.~Czekala\altaffilmark{\kipac},
 J.~Dittmann\altaffilmark{\mit},
 M.~R.~Drout\altaffilmark{\car},
 R.~J.~Foley\altaffilmark{\ucsc},
 W.~Fong\altaffilmark{\nwu},
 R.~P.~Kirshner\altaffilmark{\cfa},
 T.~Laskar\altaffilmark{\nrao,\ucb},
 C.~N.~Leibler\altaffilmark{\ucsc},
 R.~Margutti\altaffilmark{\nwu},
 D.~Milisavljevic\altaffilmark{\cfa},
 G.~Narayan\altaffilmark{\noao},
 Y.-C.~Pan\altaffilmark{\ucsc},
 A.~G.~Riess\altaffilmark{\jhu},
 K.~C.~Roth\altaffilmark{\gem},
 N.~E.~Sanders\altaffilmark{\cfa},
 D.~Scolnic\altaffilmark{\kicp},
 S.~J.~Smartt\altaffilmark{\qub},
 K.~W.~Smith\altaffilmark{\qub}, 
 K.~C.~Chambers\altaffilmark{\ifa},
 P.~W.~Draper\altaffilmark{\dur},
 H.~Flewelling\altaffilmark{\ifa},
 M.~E.~Huber\altaffilmark{\ifa},
 N.~Kaiser\altaffilmark{\ifa},
 R.~P.~Kudritzki\altaffilmark{\ifa},
 E.~A.~Magnier\altaffilmark{\ifa},
 N.~Metcalfe\altaffilmark{\dur},
 R.~J.~Wainscoat\altaffilmark{\ifa},
 C.~Waters\altaffilmark{\ifa}, and
 M.~Willman\altaffilmark{\ifa}
}

\altaffiltext{\okc}{The Oskar Klein Centre \& Department of Astronomy, Stockholm University, AlbaNova, SE-106 91 Stockholm, Sweden}
\altaffiltext{\cfa}{Harvard-Smithsonian Center for Astrophysics, 60 Garden St., Cambridge, MA 02138, USA}
\altaffiltext{\ou}{Astrophysical Institute, Department of Physics and Astronomy, 251B Clippinger Lab, Ohio University, Athens, OH 45701, USA}
\altaffiltext{\jhu}{Department of Physics and Astronomy, The Johns Hopkins
University, Baltimore, MD 21218}
\altaffiltext{\stsci}{Space Telescope Science Institute, 3700 San Martin Dr., Baltimore, MD 21218, USA}
\altaffiltext{\kipac}{Kavli Institute for Particle Astrophysics and Cosmology, Stanford University, Stanford, CA 94305, USA}
\altaffiltext{\mit}{Massachusetts Institute of Technology, 77 Massachusetts Avenue, Cambridge, Massachusetts 02138, USA}
\altaffiltext{\car}{ Carnegie Observatories, 813 Santa Barbara Street, Pasadena, CA
91101, USA 2}
\altaffiltext{\ucsc}{Department of Astronomy and Astrophysics, University of California, Santa Cruz, CA 95064, USA}
\altaffiltext{\nwu}{Center for Interdisciplinary Exploration and Research in Astrophysics (CIERA) and Department of Physics and Astronomy, Northwestern University, Evanston, IL 60208 }
\altaffiltext{\noao}{National Optical Astronomy Observatory, 950 North Cherry Avenue, Tucson, AZ 85719, USA}
\altaffiltext{\gem}{Gemini Observatory, 670 North Aohoku Place, Hilo, HI 96720, USA}
\altaffiltext{\nrao}{National Radio Astronomy Observatory, 520 Edgemont Road, Charlottesville, VA 22903, USA}
\altaffiltext{\ucb}{Department of Astronomy, University of California, 501 Campbell Hall, Berkeley, CA 94720-3411, USA}
\altaffiltext{\kicp}{Kavli Institute for Cosmological Physics, University of Chicago, Chicago, IL 60637, USA}
\altaffiltext{\qub}{Astrophysics Research Centre, School of Mathematics and Physics, Queen’s University Belfast, Belfast BT7 1NN, UK}
\altaffiltext{\ifa}{Institute for Astronomy, University of Hawaii, 2680 Woodlawn Drive, Honolulu, HI 96822, USA}
\altaffiltext{\dur}{Department of Physics, Durham University, South Road, Durham DH1 3LE, UK}

\begin{abstract}
We present light curves and classification spectra of 17 hydrogen-poor superluminous supernovae (SLSNe) from the Pan-STARRS1 Medium Deep Survey (PS1~MDS). Our sample contains all objects from the PS1~MDS sample with spectroscopic classification that are similar to either of the prototypes SN\,2005ap or SN\,2007bi, without an explicit limit on luminosity. With a redshift range $0.3 < z < 1.6$, PS1~MDS is the first SLSN sample primarily probing the high-redshift population; our multi-filter PS1 light curves probe the rest-frame UV emission, and hence the peak of the spectral energy distribution. We measure the temperature evolution and construct bolometric light curves, and find peak luminosities of $(0.5-5) \times 10^{44}$~erg~s$^{-1}$ and lower limits on the total radiated energies of $(0.3-2) \times 10^{51}$~erg. The light curve shapes are diverse, with both rise- and decline times spanning a factor of $\sim 5$, and several examples of double-peaked light curves. When correcting for the flux-limited nature of our survey, we find a median peak luminosity at 4000~\AA\, of $M_{\rm 4000} = -21.1~{\rm mag}$, and a spread of $\sigma = 0.7~{\rm mag}$.
\end{abstract}


\section{Introduction}
Superluminous supernovae (SLSNe) are a rare class of supernovae (SNe) discovered in galaxy-untargeted transient surveys over the past decade. They are characterized by peak luminosities of $10-100$ times those of normal core-collapse and Type Ia SNe, and are significantly rarer ($\sim 0.01\%$ of the core-collapse SN rate; \citealt{qya+13,msr+14,pss+17}). With total radiated energies of order $10^{51}~{\rm erg}$, their light curves are difficult to explain with conventional SN energy sources, and as a result this class has garnered significant attention.

SLSNe can be divided into two spectroscopic subclasses, based on the presence or absence of hydrogen in the spectrum. The majority of H-rich SLSNe (often dubbed SLSN-II) show narrow Balmer lines similar to Type IIn SNe, likely powered by interaction with a dense circumstellar medium (CSM) \citep[e.g.][]{ock+07,slf+07, scs+10, cwv+11, ddm+11b, rfg+11}. However, there are also examples of SLSN-II without clear spectroscopic interaction signatures \citep{ghg+09,mcp+09,isg+16}, as well as objects classified as SLSN-I based on their peak spectra but show hydrogen features at late times \citep{yqo+15,ylp+17}.  

For SLSNe without hydrogen signatures in their spectra (H-poor SLSNe, or SLSN-I), the power source is still debated. CSM interaction has also been proposed as a mechanism for this subclass, but would require an extreme mass-loss history in order to reproduce the observed light curves: several $M_{\odot}$ of H-poor material lost in the last $\sim$year before explosion \citep{ci11,gb12, cw12b, mbt+13}. The lack of narrow lines seen in the spectra at any epoch is also a puzzle if CSM interaction is the power source. Alternative explanations include a central-engine model, such as the spin-down of a newborn magnetar energizing the ejecta over timescales of weeks \citep{woo10,kb10,dhw+12,mmk+15,wwd+15}. This model has gained popularity thanks to its ability to explain a wide variety of SLSN light curves \citep[e.g.][]{ccs+11,isj+13,lcb+13,lcb+16,nsj+13,nbs+16,nbm+17}, though a smoking-gun signature of the magnetar engine, such as X-ray break-out \citep{mvh+14} remains elusive \citep{mcm+17}. Finally, the slowest-evolving H-poor SLSNe have been proposed to be pair-instability supernovae (PISNe) powered by the radioactive decay of several solar masses of $^{56}$Ni \citep{brs67,gmo+09}, and sometimes referred to as ``SLSN-R'' \citep{gal12}. This interpretation is controversial, however, as models like magnetar spin-down can also explain these SLSNe \citep{ysv+10,dhw+12,nsj+13,lcb+16}. The bolometric luminosity of the these events tend to fall below that expected from fully trapped $^{56}$Co decay \citep{inc+17,csj+14}, and the emission line strengths of $\alpha$-processed elements (oxygen and magnesium)
indicate ejecta masses of 10-30~${\rm M}_{\odot}$ \citep{jsi+17}. Neither of these observations sit comfortably with  pair-instability 
model predictions. Therefore, whether ``slowly-evolving'' H-poor SLSNe represent a separate subclass, and if so, what physical mechanism is responsible, is still an open question.

Beyond their energy sources, SLSNe have garnered significant attention as potential probes of the high-redshift universe. Both due to their overall high luminosities and because their spectral energy distributions (SEDs) peak in the ultraviolet (UV), SLSNe are observable to much higher redshifts than ordinary SNe, making them excellent targets for high-redshift SN searches. Currently, spectroscopically classified SLSNe have been found out to redshifts $z \simeq 2$ \citep{gdp+16,pfs+17}, and candidate SLSNe out to redshifts $z \sim 4-6$ \citep{csg+12,mac+17}. Studies of literature samples of SLSNe have suggested that the scatter in SLSN-I luminosities is intrinsically low and can be further improved by considering correlations with colors and decline rates \citep{is14,pds+15}, leading to increased interest in the potential use of SLSNe as standardizable candles \citep{wwm15,snb+16}. Beyond potential cosmology applications, high-redshift SLSNe also offer a probe of studying high-redshift galaxies \citep{bcl+12,vsg+14}.

Since SLSNe are rare, previous studies have largely focused on individual events, or combined data from the literature from many different surveys. Here, we present the full sample of H-poor SLSNe discovered in the Pan-STARRS1 Medium Deep Survey (PS1~MDS) over its four years of operation, comprising of 17 events over a redshift range $0.3 < z < 1.6$. This is the first single-survey compilation study that covers primarily the high-redshift population (see \citealt{dgr+17} for a compilation of the generally lower-redshift SLSN-I sample from the Palomar Transient Factory).
We describe the survey parameters, our selection criteria for designating a transient as a SLSN, and present the classification spectra and observed light curves in Section~\ref{sec:comp_data}. Inferred physical properties, such as temperature evolution, expansion velocities, bolometric light curves and total radiated energies are presented in Section~\ref{sec:comp_phys}. We explore the light curve shapes, including rise and decline times, and double peaked light curves, in Section~\ref{sec:lcshape}, and model fits to some of our best-sampled light curves that have not been previously published in Section~\ref{sec:magnetar}.
Implications of our findings are discussed in Section~\ref{sec:comp_disc}, and summarized in Section~\ref{sec:comp_conc}. Throughout this paper, we assume a flat $\Lambda$CDM cosmology with $\Omega_{\rm M} = 0.27$, $\Omega_{\Lambda} = 0.73$ and $H_0 = 70$~km~s$^{-1}$ \citep{ksd+11}.

\section{The PS1~MDS SLSN Sample}
\label{sec:comp_data}
\subsection{Pan-STARRS1 Medium Deep Survey}

The PS1 telescope on Haleakala is a high-etendue wide-field survey
instrument with a 1.8-m diameter primary mirror and a $3.3^\circ$
diameter field of view imaged by an array of sixty $4800\times 4800$
pixel detectors, with a pixel scale of $0.258''$
\citep{PS1_system,PS1_GPCA}.   \citet{tsl+12} describes the photometric system and broadband filters in detail. 

The Pan-STARRS1 system and its surveys are fully described in \citet{cmm+16}. The stacked 3$\pi$ survey data are publicly available from the Space Telescope Science Institute archive \footnote{\url{http://panstarrs.stsci.edu/}}. This paper describes the data taken from the Pan-STARRS1 Medium deep Survey (MDS) designed by the Pan-STARRS1 Science Consortium (PS1SC). The Pan-STARRS1 Medium Deep Survey (PS1~MDS) operated from late 2009 to early 2014. PS1~MDS consists of 10 fields (each with a single PS1 imager footprint). The fields were observed in \gps\rps\ips\zps\ with a typical cadence of 3~d in each filter, to a typical nightly depth of $\sim 23.3$ mag ($5\sigma$); \yps\ was used near full moon with a typical depth of $\sim 21.7$ mag (AB mags are used throughout this paper).
The standard reduction, astrometric solution, and stacking of the nightly images were performed by the Pan-STARRS1 Image Processing Pipeline (IPP) system on a on a computer cluster originally 
based at the Maui High Performance Computer Center. The processing steps 
to reduce and stack the data are described in \citet{mcf+16} and in \citet{wmp+16}, 
while the steps for astrometric calibration are in \citet{msf+16} For the transients search, the nightly MDS stacks were transferred
to the Harvard FAS Research Computing cluster, where they were
processed through a frame subtraction analysis using the {\tt photpipe}
pipeline developed for the SuperMACHO and ESSENCE surveys \citep{rsb+05, gsc+07, mpr+07,rsf+14}. An additional set of difference images were produced by the IPP in Hawaii, and 
the catalogues of the detections were ingested into a database at Queen's
University Belfast (see \citealt{msr+14}). Cross-matches between the two 
end to end pipelines were made, to mitigate loss of transients through either. 

A subset of targets was selected for spectroscopic follow-up, using the Blue Channel spectrograph on the 6.5-m MMT telescope \citep{swf89}, the Gemini Multi-Object Spectrograph (GMOS; \citealt{hja+04}) on the 8-m Gemini telescopes, and the Low Dispersion Survey Spectrograph (LDSS3) and Inamori-Magellan Areal Camera and Spectrograph (IMACS; \citealt{dhb+06}) on the 6.5-m Magellan telescopes. The SLSNe were generally targeted for spectroscopy based on a combination of blue observed color, long observed rise time, and being several magnitudes brighter than any apparent host in the PS1 deep stacks -- for more details, see Section~5.1 of \citet{lcb+14} which discusses both the selection and possible biases introduced. We note that the combination of a modest survey area and deep photometry provides sensitivity primarily to SLSNe at higher redshifts: the sample spans $ 0.3 \lesssim z \lesssim 1.6$. Table~\ref{tab:targets} lists the full sample.

\subsection{Classification Spectra}

As we are interested in the true luminosity range of SLSNe, we do not include a luminosity threshold in our definition, and instead adopt a spectroscopy-based selection. We define our sample of SLSNe as SNe that are spectroscopically similar to either of the prototypes SN\,2005ap/SCP06F6 (2005ap-like) or to SN\,2007bi (2007bi-like). While this is reminiscent of the division by \citet{gal12} into ``SLSN-I'' and ``SLSN-R'', we do not include any light curve information, or intend to imply anything regarding the power source by making this distinction; we simply wish to include all kinds of H-poor SLSNe. Indeed, there are examples of objects (e.g. PS1-11ap and PTF12dam; \citealt{nsj+13,msk+14}) that resembled SN\,2005ap near peak but developed features similar to SN\,2007bi on the decline; it has been suggested that the differences mainly arise due to temperature effects \citep{ngb17}. Here, we use the spectrum taken closest to peak light for classification, to make the selection as uniform as possible. Using peak spectra also minimizes confusion with SN Ib/c, as SLSN-I on the decline often develop features similar to SN Ib/c at peak, as the ejecta cool and the photosphere reaches comparable temperatures \citep[e.g.][]{psb+10,msp+16}. With these criteria, we find 16 2005ap-like objects in the PS1~MDS spectroscopic sample, and one 2007bi-like (PS1-14bj, discussed in detail in \citealt{lcb+16}). All classification spectra are shown in Figure~\ref{fig:spec_05ap}, and the details of spectroscopic observations (if previously unpublished) are listed in Table~\ref{tab:spectra}. 

In practice, given the redshifts of our objects the features most commonly used for classification of the 2005ap-like objects was the series of broad UV features bluewards of 2800~\AA\,, marked on the spectrum of SCP06F6 in Figure~\ref{fig:spec_05ap}. Despite many objects having limited wavelength coverage, all but two of our spectra go sufficiently blue to cover at least the broad \ion{Mg}{2} feature, which is the reddest of the series. In the optical, the series of characteristic \ion{O}{2} features (marked on the spectrum of PTF09cnd) are comparatively shallower, and typically stronger during the rise of the light curve than at peak light. These features are convincingly present at peak in PS1-11ap, PS1-10bzj and PS1-13gt, though PS1-13gt is the only objects where the classification is based on the \ion{O}{2} absorption as opposed to the UV features. The fact that the redshift is unambiguously known from narrow host galaxy features in the majority of cases also aids classification, particularly in cases where there are just a few discernible features in the spectrum and/or the wavelength coverage is limited.

A few objects have been discussed in the literature as SLSNe from PS1~MDS but are not included in our sample here. One such object is PS1-12zn, which was included in the sample of H-poor SLSNe in the host galaxy study of \citet{lcb+14}. Although we do not detect H lines in its spectrum, and its luminosity places it firmly in the SLSN category, the spectrum shows a featureless blue continuum, lacking both the broad UV features and the \ion{O}{2} features we use here as our spectroscopic criteria. As our spectrum does not cover H$\alpha$, we cannot rule out that this object was a H-rich SLSN, and we therefore do not include it in our spectroscopically selected sample here.
We also exclude PS1-10afx, presented as a possible SLSN in \citet{cbr+13}, as the discovery of a second galaxy along the line of sight has revealed this object to be a lensed SN Ia \citep{qwo+13,qom+14}, and its spectrum is indeed better matched to a normal SN Ia than to SN\,2005ap or SN\,2007bi.

All but three objects in our sample have narrow-line host galaxy redshifts from either [\ion{O}{2}] $\lambda$3727 emission or \ion{Mg}{2}\,$\lambda\lambda$2796,2803 absorption lines. For the three objects without any host galaxy absorption or emission lines, PS1-12cil, PS1-10ahf and PS1-13or, we instead determine the redshift from the supernova spectra. The higher-redshift objects were matched to the series of strong UV features seen in SCP06F6 and PS1-10ky \citep{bdt+09,ccs+11}. Owing to its lower redshift, only the first of these features is detected in PS1-12cil, but post-peak spectra of this object (Chornock et al. 2017, in preparation) develop features similar to SN Ib/c post-peak as the ejecta cool (similar to other SLSN-I). We use these later spectra, cross-correlated to SN Ib/c templates using SNID \citep{snid} to determine the redshift of this object. We caution that redshifts derived from supernova features are degenerate with the expansion velocity, and therefore less precise than the narrow-line redshifts reported for the rest of the sample, and we only report the redshift to two decimal places for these objects.

\begin{figure*}
\begin{center}
\includegraphics[width=7in]{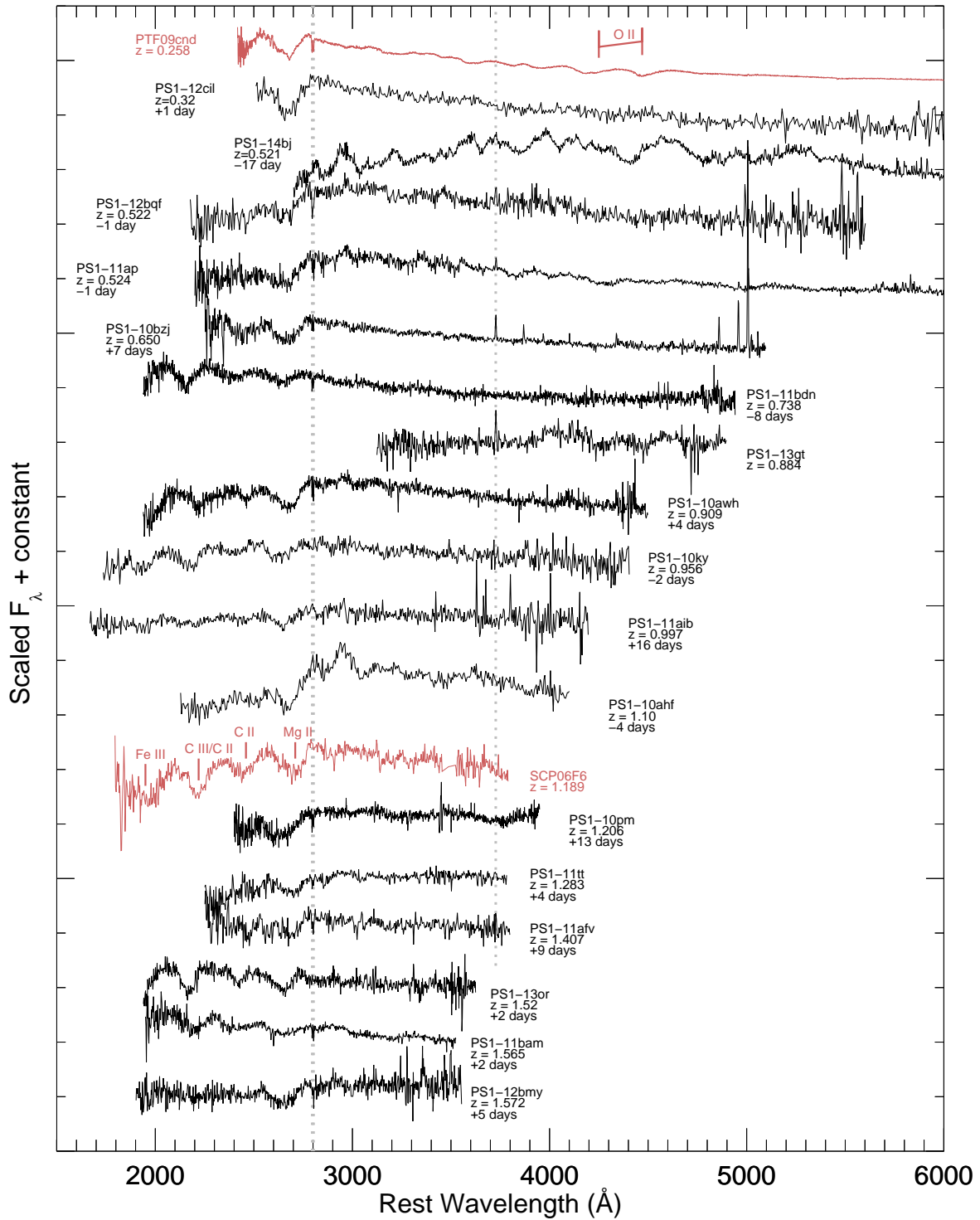}
\caption{Classification spectra of the 17 SLSNe in our sample, taken as close to peak light as possible (actual phase indicated for each spectrum). The dashed gray lines mark the location of the Mg II $\lambda\lambda$2796,2803 doublet and the [O II]$\lambda$3727 emission line, which were used to determine the redshift for most of these objects. Spectra have been arbitrarily scaled and binned for display purposes. With the exception of PS1-14bj, which is notably redder and shows more features, all objects are spectroscopically similar to SN\,2005ap. PTF09cnd (pre-peak; \citealt{qkk+11}) and SCP06F6 (at peak; \citealt{bdt+09}) are shown in red as comparisons.
\label{fig:spec_05ap}}
\end{center}
\end{figure*}

\subsection{Light Curves}
Thanks to the multi-band data from PS1~MDS, \gps\rps\ips\zps\ light curves are available for all objects. Most objects are undetected in the shallower \yps\ band, and we find that the upper limits do not provide meaningful constraints -- we therefore only report \yps\ photometry for the three objects that are actually detected: PS1-12cil, PS1-12bqf and PS1-11ap.  The final photometric pipeline is described in \citet{sjr+17}.
In addition to the PS1 photometry, some objects have additional follow-up imaging acquired with GMOS, LDSS and IMACS; we reduced these images and extracted magnitudes by aperture photometry using standard routines in IRAF. PS1-11bdn was also observed with the \textit{Swift} Ultra-Violet/Optical Telescope (UVOT); magnitudes in a 3\arcsec aperture were extracted following the procedure in \citet{bhi+09}. All photometry is listed in Table~\ref{tab:phot}. The light curves of the 17 objects in our sample are shown in Figure~\ref{fig:lcgold}. Due to the large redshift range of our sample, the effective wavelengths of each filter varies significantly; Table~\ref{tab:leff} lists these effective wavelengths at the redshift of each supernova. 

Note the long observed timescales in many cases, due both to intrinsically longer timescales of SLSNe, as well as time dilation. The long timescales also mean that depending on when an object was discovered during an observing season, we may not have a complete light curve. Particularly among the higher-redshift objects, we tend to sample either the rise or the decline, although we do observe either a turnover or flattening of the flux suggesting we are capturing the peak in most cases. Exceptions to this, where the time of peak is uncertain, include PS1-13gt (which is declining in all filters), PS1-11bdn (which has a very sparsely sampled light curve), and to a lesser extent PS1-10ahf and PS1-13or.

\begin{figure*}
\begin{center}
\includegraphics[width=7in]{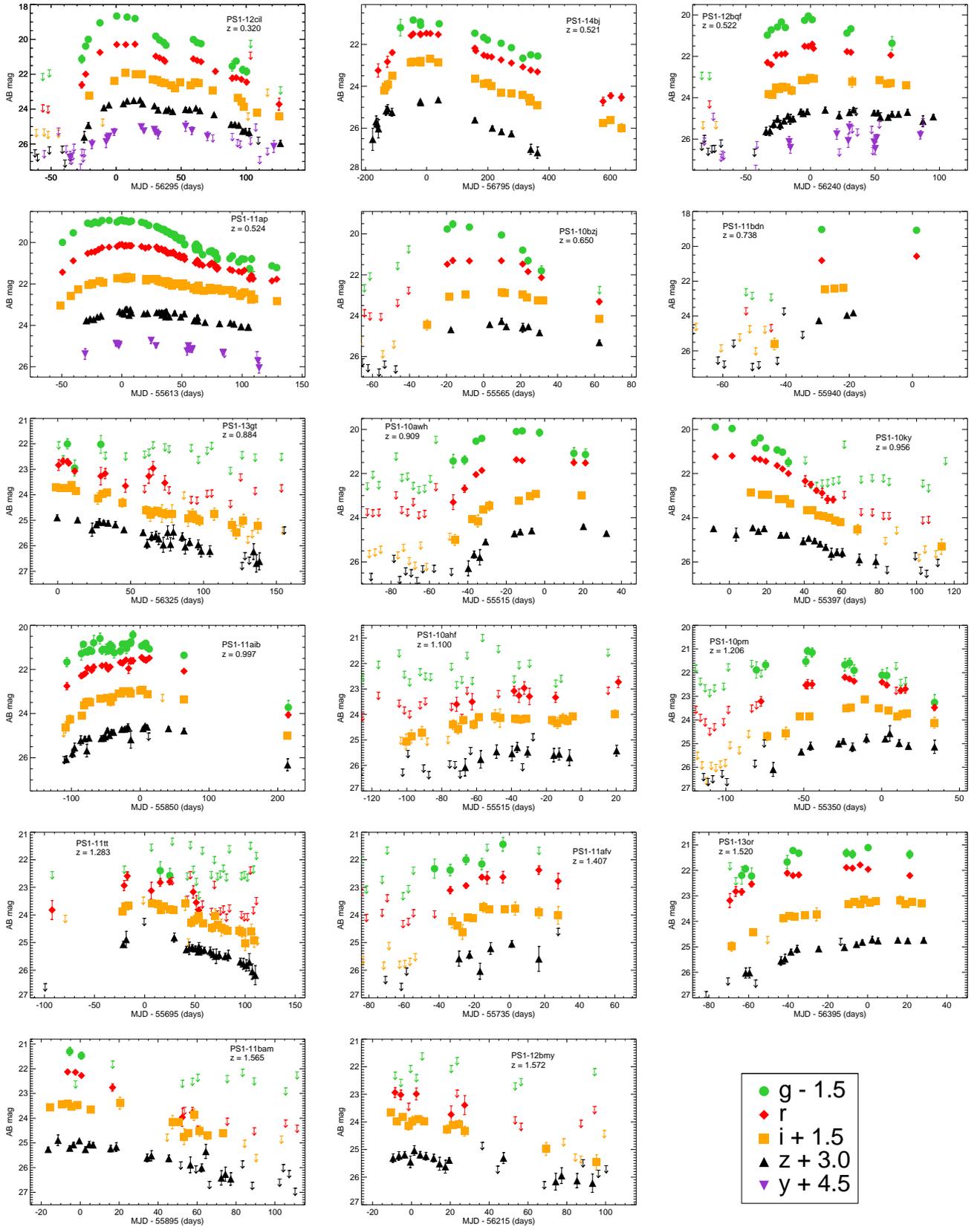}
\caption{Multiband light curves of all 17 events in our H-poor SLSN sample, sorted in order of redshift. Filters are offset by 1.5~mag for clarity, as indicated by the legend in the bottom right panel. \yps\ light curves are included only for the objects that are detected in this shallower filter. Table~\ref{tab:leff} lists the effective rest wavelengths of each filter for each supernova.
\label{fig:lcgold}}
\end{center}
\end{figure*}

\section{Derived Physical Parameters}
\label{sec:comp_phys}

\subsection{Color and Temperature Evolution}
\label{sec:color}

The multiband nature and large redshift range of the PS1~MDS SLSN sample allow us to probe the light curves and colors of SLSNe in the rest-frame UV. The observed colors at peak are shown in Figure~\ref{fig:color}. The strongest trend with redshift is seen in \gps$-$\rps, as the peak of the SED moves through the observed \gps band; this is also illustrated in Figure~\ref{fig:filter_redshift}, which shows the PS1 filter curves at different redshifts compared with typical SLSN-I spectra. \gps$-$\rps also shows the largest scatter at a given redshift, reflecting the corresponding spread in UV luminosities. Such a spread is also seen among well-studied low-redshift SLSNe with good UV coverage; see e.g. the very UV luminous SLSN Gaia16apd \citep{nbm+17,yqg+17,kbm+17}. This illustrates a challenge in using colors in identifying high-redshift SLSNe. \rps $-$ \ips and \ips$-$\zps are flatter with redshift and have less scatter. 

We correct the photometry for foreground extinction following \citet{sf11}, but are not able to correct for the (unknown) host galaxy extinction. This could also be contributing to the spread in observed colors, although the host galaxies of most of the SLSNe in this sample were studied in \citet{lcb+14} and found to have little inferred dust extinction; the same result is found in other studies and appears to be true for SLSN host galaxies in general \citep[e.g.][]{lsk+15,pqy+16}. Therefore, we do not generally expect a large contribution from the host galaxies; this is also supported by the low average extinction found in the modeling by \citet{ngb17}. In individual cases, host galaxy reddening may still be important, however. Another uncertainty is reddening by dust associated with circumstellar material lost by the progenitor star -- some models of SLSNe predict eruptive mass loss prior to explosion (e.g., \citealt{woo17}), and late-time interaction signatures in some H-poor SLSNe also supports the idea of a complex circumstellar environment \citep{yqo+15,ylp+17}. Depending on the distance to the CSM, dust is likely destroyed by the supernova radiation, however, so this may not be an important effect.

One SLSN in our sample with signs of possible reddening is PS1-13gt, which shows a comparatively red continuum despite also showing the characteristic \ion{O}{2} features in its spectrum that require high temperatures (Figure~\ref{fig:spec_05ap}, e.g.~\citealt{msp+16}). This suggests that the temperature is higher than one would infer from the shape of the continuum. When correcting the spectrum to rest-frame wavelengths and dereddening by E($B-V$)$\simeq 0.3~{\rm mag}$, the spectrum of PS1-13gt is an excellent match to PTF09cnd \citep{qkk+11}. This SLSN is also one of the faintest found in the sample, which supports the possibility of higher extinction.

\begin{figure}
\centering
\includegraphics[width=3.4in]{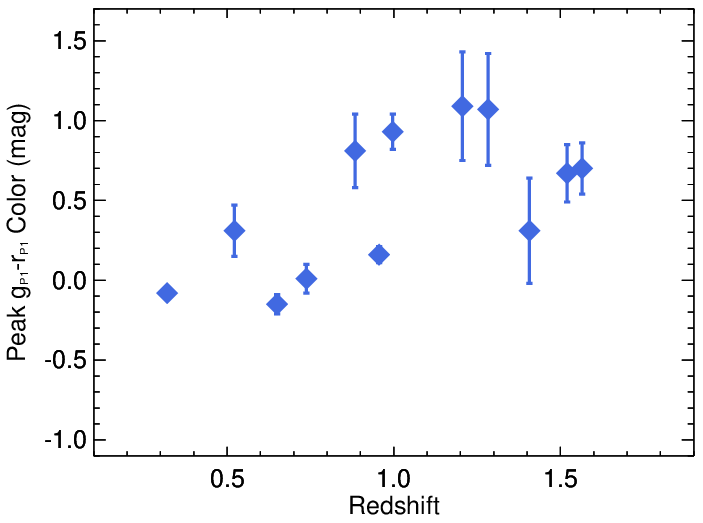} \\
\includegraphics[width=3.4in]{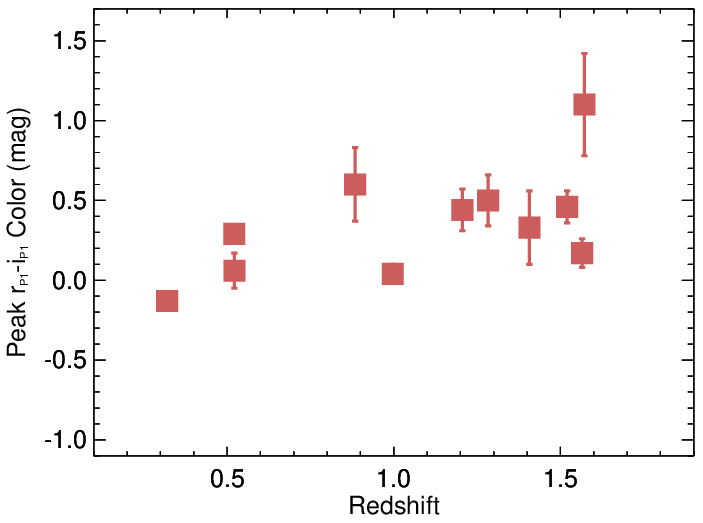} \\
\includegraphics[width=3.4in]{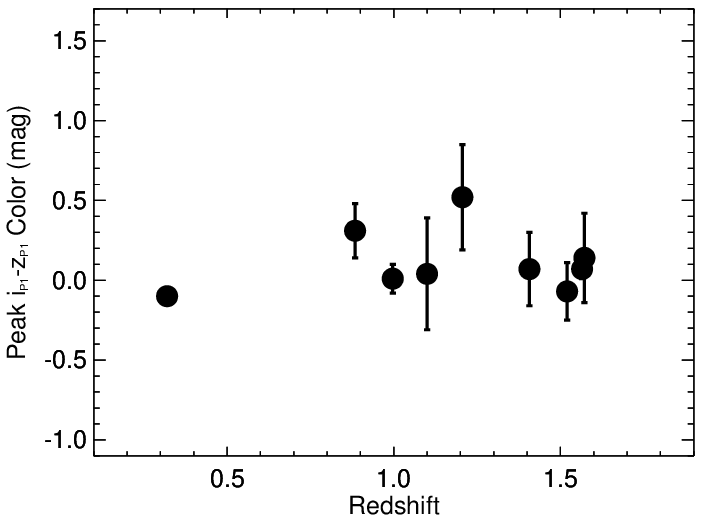} 
\caption{Observed color at peak as a function of redshift. \gps$-$\rps\ in particular shows appreciable scatter even over a small range in redshift, reflecting the spread in UV luminosities in the sample. 
}
\label{fig:color}
\end{figure}

\begin{figure}
    \centering
    \includegraphics[width=3.4in]{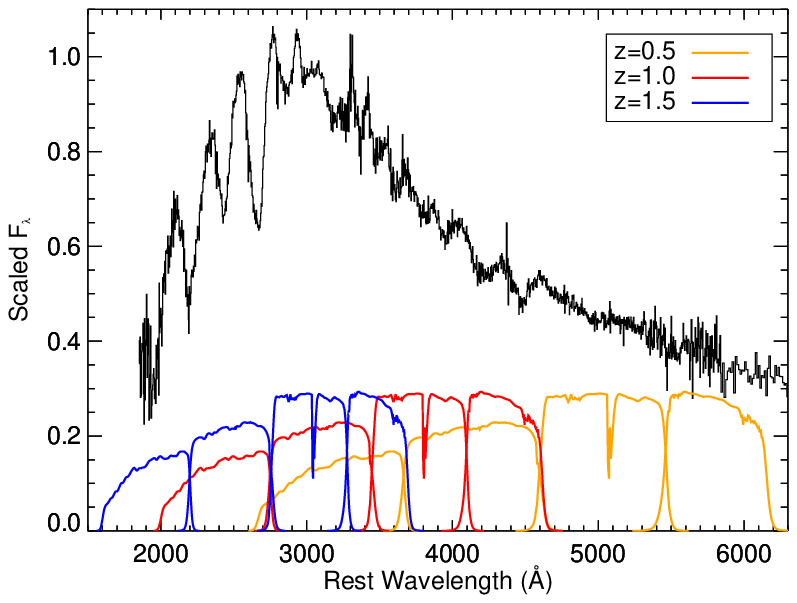}
    \caption{Spectrum of the typical SLSN-I iPTF13ajg \citep{vsg+14}, with the \gps\rps\ips\zps filter curves at three different effective redshifts overplotted, illustrating the effect of redshift on observed color (Figure~\ref{fig:color}). For example, past redshift $z \simeq 1$, \gps and \rps sample the part of the spectrum with strong UV absorption, contributing to both the larger scatter and redder colors seen in \gps$-$\rps at higher redshifts. \ips$-$\zps, in contrast, probe the optical with relatively weak absorptions over most of the redshift range covered, and show comparatively little evolution.}
    \label{fig:filter_redshift}
\end{figure}

We measure temperature as a function of time from the light curves by fitting blackbody curves to the observed photometry. Typically, PS1 observed \gps and \rps on the same night, so we generally use \rps as the baseline for these calculations. If there is photometry from the other bands from the same night or $\pm$ 1 day, we use those measurements without corrections. If not, we use a polynomial fit to the light curve in that filter and interpolate to the date of the \rps observation. We only fit SEDs to epochs where the object was observed in at least 3 filters. Figure~\ref{fig:bbtemp} shows the resulting blackbody temperatures derived from the photometry.

Early measurements in particular are noisy, because the peak of the blackbody can be bluewards of the observed bands, even for the high-redshift PS1~MDS sample. To the extent that we can measure it, we find that the color temperatures prior to peak are either constant or slowly cooling, with temperatures in the range $10,000 - 25,000~{\rm K}$. This highlights the need for UV follow-up of SLSNe, particularly at early epochs. Post-peak, the color temperatures decrease as the SN ejecta expand and cool, and also seem to plateau around $6,000-7,000~{\rm K}$.

PS1-14bj deviates from this general trend, having redder colors and cooler temperatures over the entire observed time period, and an overall flat color evolution. PS1-13gt shows the reddest color temperature at peak, which would be consistent with this supernova being reddened by dust as discussed above.

\begin{figure*}
\begin{center}
\includegraphics{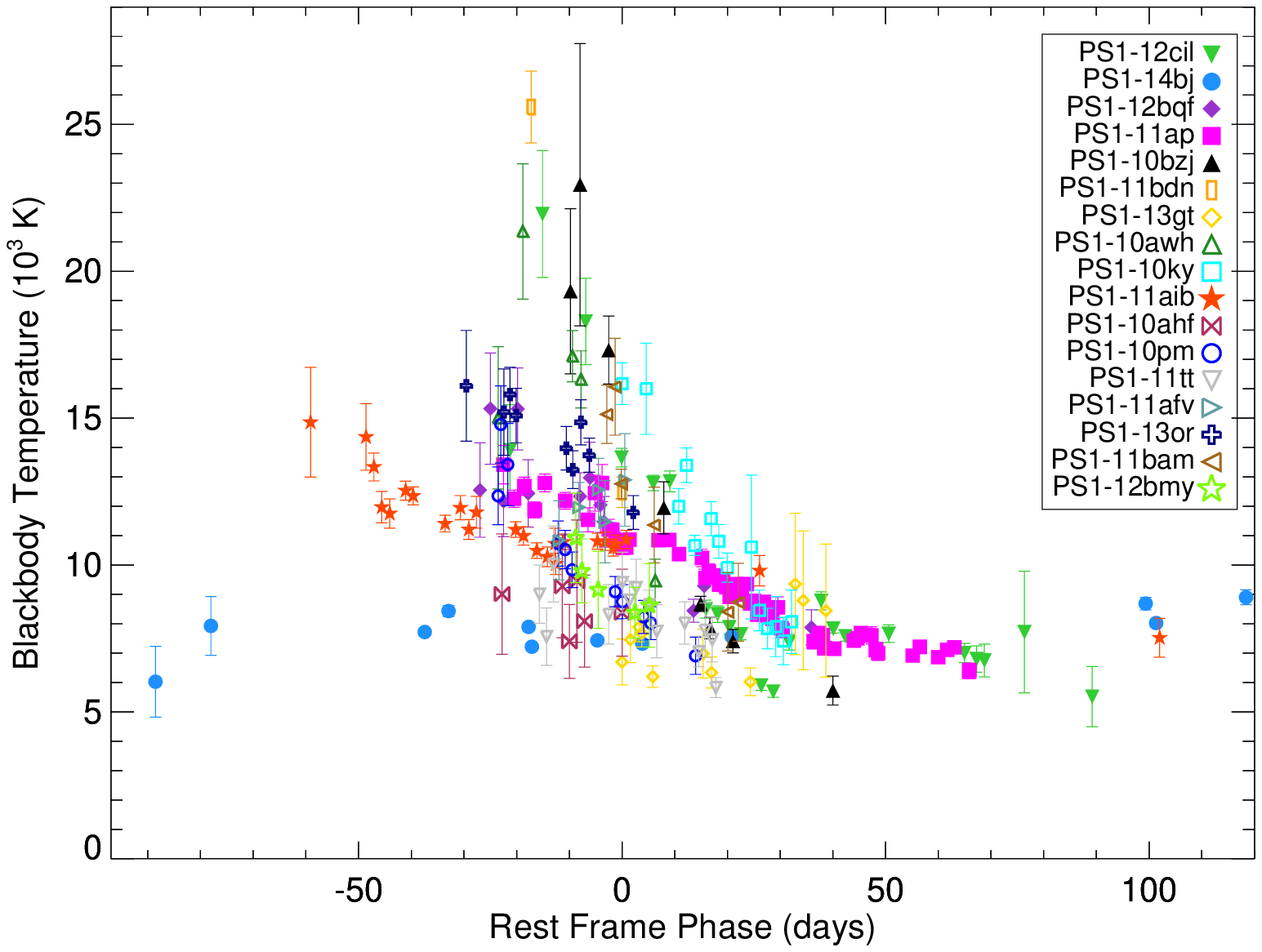}
\caption{Color temperature as a function of phase, measured by fitting a blackbody to the photometry at each epoch. Prior to peak, the 2005ap-like SLSNe show hot color temperatures around $10,000-25,000$~K, and cool over a timescale of $20-50$~days after peak light. PS1-14bj, our only 2007bi-like SLSN, shows color temperatures of $6000-8000$~K over its entire evolution.
\label{fig:bbtemp}}
\end{center}
\end{figure*}

\subsection{Expansion Velocity}
We measure velocities from the spectra by fitting Gaussians to the absorption features and determining the locations of the minima. The identification of the strong UV features is debated -- \citet{qkk+11} identified them with C II, Si III and Mg II, whereas \citet{hkl+13} favors Fe III, C II/III and Mg II; see also \citet{msp+16}. We do not attempt modeling of the spectra given the spread in quality and wavelength coverage for our objects. However, in all but one of our SLSNe that are classified as 2005ap-like our spectra cover the broad Mg II feature, and we use this to estimate the velocity at peak, and calculate the associated velocity from the blueshift relative to the narrow Mg II lines from the host galaxy. Table~\ref{tab:results} lists the expansion velocities derived in this fashion. They range from $10,000$ to $18,000$~km~s$^{-1}$, with typical values of about $15,000$~km~s$^{-1}$. This is similar to what has been seen in other SLSNe around peak light (e.g., \citealt{qkk+11,isj+13,nsj+14,lmb17}).

\subsection{Bolometric Light Curves and Total Radiated Energies}
To construct bolometric light curves, we start with the observed photometry at each multi-filter epoch (i.e., flux and effective wavelength for each filter), and sum up the observed flux using trapezoidal integration. We linearly extrapolate the flux from the effective wavelength to the blue edge of the bluest filter (typically \gps) and red edge of the reddest filter (typically \zps). To be explicit, for a series of fluxes $\{f_0, f_1, ..., f_n\}$ with corresponding effective wavelengths $\{\lambda_0, \lambda_1, ..., \lambda_n\}$, and the blue edge of the bluest filter $\lambda_b$, red edge of reddest filter $\lambda_r$, we calculate

\begin{flalign}
&L_{\rm trap} =\sum_{k=1}^{n}\frac{f_k + f_{k-1}}{2} \left(\lambda_k - \lambda_{k-1}\right) \\ \nonumber
& + f_0 \left(\lambda_0 - \lambda_b\right) + f_n\left(\lambda_r - \lambda_n\right) \\ \nonumber 
&  -  \frac{f_1-f_0}{2}\frac{\left(\lambda_0-\lambda_b\right)^2}{\left(\lambda_1-\lambda_0\right)} + \frac{f_n-f_{n-1}}{2}\frac{\left(\lambda_r-\lambda_n\right)^2}{\left(\lambda_n-\lambda_{n-1}\right)}. 
\end{flalign}

\noindent Since this only takes into account the flux in the observed bands, it is a strict lower limit on the emitted flux. Given the large redshift range of our sample, the rest-frame wavelengths covered in this estimate also varies considerably; see Table~\ref{tab:leff} for the actual rest frame wavelengths covered at the redshift of each object.

For a better estimate of the bolometric luminosities, we add a correction to the observed flux based on the estimated blackbody temperatures. While the spectrum clearly deviates from a blackbody at UV wavelengths (bluewards of the observed bands; Figure~\ref{fig:spec_05ap}), it is reasonably well approximated by a blackbody at redder wavelengths. We therefore integrate a blackbody curve redwards of the observed bands, with the observed color temperature and scaled to match the flux in the reddest observed filter, and add this to the observed flux. The size of this correction is small (10-20\%) at early times, but can be substantial at later times as the SNe cool and the blackbody peak shifts to the red. Similarly, the correction is larger for the higher-redshift objects, as the observed filters cover bluer rest-frame wavelengths. We explore the luminosity function in a standardized bandpass in Section~\ref{sec:comp_disc}.

Figure~\ref{fig:bollc} shows the pseudo-bolometric light curves calculated in this fashion. While only a handful of light curves are well sampled both before and after peak, the diversity in light curve shapes is still apparent. We explore this in a more quantitative way in Section~\ref{sec:lcshape}.

\begin{figure*}
\begin{center}
\includegraphics{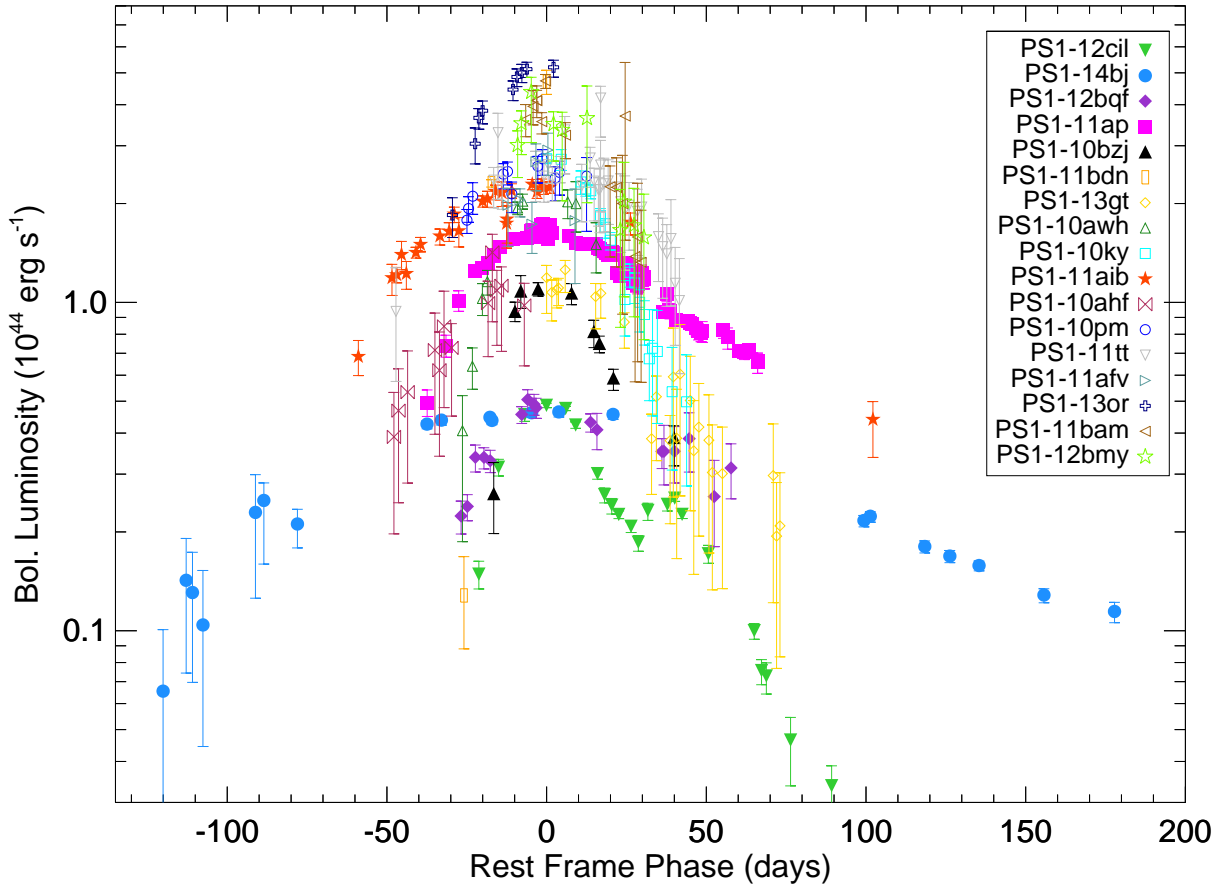}
\caption{Pseudo-bolometric light curves, created by summing up the observed flux and adding a blackbody tail in the red. Where there is insufficient color information at the very beginning or end of a light curve, points have been plotted assuming a constant bolometric correction. We use these light curves to measure peak bolometric luminosities, total radiated energies, and rise- and decay timescales. 
\label{fig:bollc} }
\end{center}
\end{figure*}

Figure~\ref{fig:peaklum} shows the peak luminosities in erg~s$^{-1}$, plotted as a function of redshift. The typical uncertainties, which can also be gleaned from Figure~\ref{fig:bollc}, are of order 10\% -- we caution, however, that systematic uncertainties due to not capturing the full bolometric flux likely dominate the statistical uncertainty. We see a clear spread of luminosities at redshifts $z \lesssim 1$, where we are sensitive also to lower-luminosity objects. This illustrates the need to take into account the impact of survey and follow-up limits on the resulting luminosity distribution of SLSNe. At redshifts $z \gtrsim 1$, we are dominated by the higher-luminosity objects, as one might expect due to Malmquist bias. Note that the low- and high-redshift luminosities are not directly comparable since our pseudo-bolometric light curves capture more of the UV light at higher redshifts -- given that the overall trend towards higher luminosities at higher redshifts holds also when comparing $K$-corrected peak magnitudes (Section~\ref{sec:comp_disc}, Figure~\ref{fig:absmag_vs_z}), this is unlikely to be a dominant effect, however. With the exception of the lowest-luminosity objects PS1-12cil, PS1-12bqf and PS1-14bj, all of the PS1 H-poor SLSNe peak at $(1-5) \times 10^{44}~{\rm erg s}^{-1}$.

\begin{figure}
\begin{center}
\includegraphics{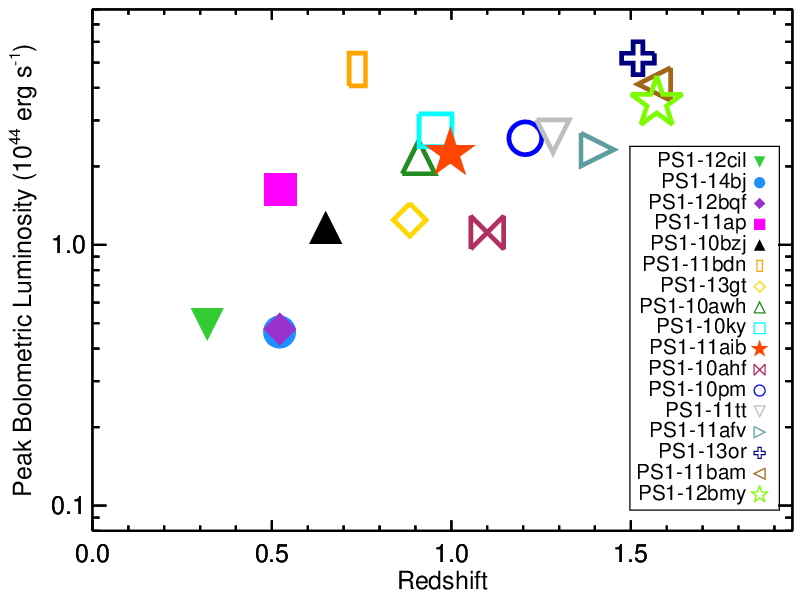}
\caption{Luminosities at peak, as measured from our pseudo-bolometric light curves (Figure~\ref{fig:bollc}). Higher redshift objects have more of the UV flux included as the bolometric estimates, so the numbers at high- and low redshift are not directly comparable; the numbers at low redshift alone show that the peak luminosities of SLSNe can vary by almost an order of magnitude, however. The low scatter at the high redshift end is due to the limitations of spectroscopic follow-up: objects like PS1-12bqf would be too faint to classify at these redshifts. 
\label{fig:peaklum} }
\end{center}
\end{figure}

We determine a lower limit on the total radiated energy by integrating the estimated bolometric light curves; the results are plotted in Figure~\ref{fig:erad}. Filled symbols correspond to objects were we sample both the rise and the decline; the results span close to an order of magnitude. Both light curve shape and overall luminosity contribute to this scatter -- while PS1-14bj and PS1-12bqf are the lowest-luminosity objects in the sample, the total radiated energy of PS1-14bj is comparable to the higher-luminosity objects, thanks to the exceptionally broad light curve. By contrast, the radiated energies of PS1-12cil and PS1-12bqf are the lowest of all in the sample, despite incomplete light curves for several of the other SLSNe. 

\begin{figure}
\begin{center}
\includegraphics{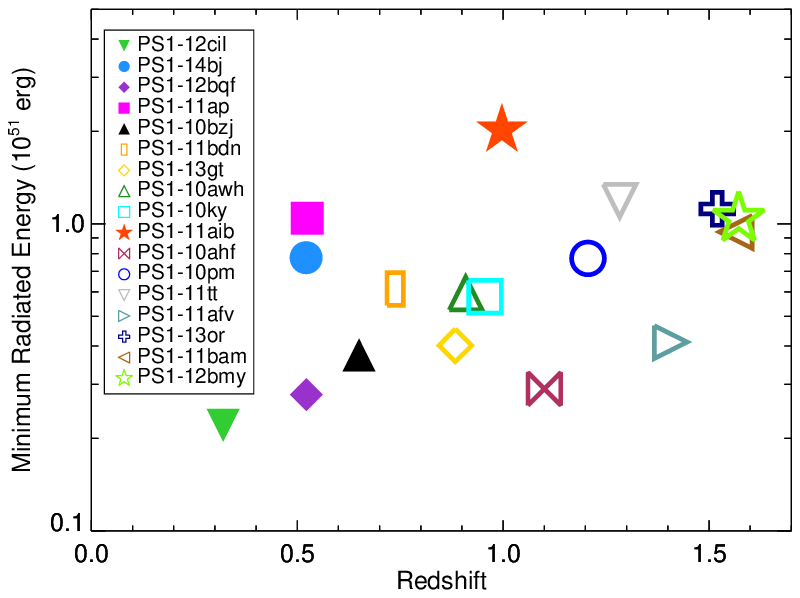}
\caption{Lower limits on the total radiated energies, again as a function of redshift, measured by integrating the pseudo-bolometric light curves (Figure~\ref{fig:bollc}). Numbers plotted here are lower limits, both because we do not in general cover the entire light curve, and because we are not accounting for the flux bluewards of the observed bandpasses. 
\label{fig:erad} }
\end{center}
\end{figure}

\section{Light Curve Shapes}
\label{sec:lcshape}

\subsection{Rise- and Decline Timescales}
We measure the time of peak, and the rise and decline times by fitting low-order polynomials to our pseudobolometric light curves. For estimates of the rise and decline times, we follow \citet{nsj+15b} and define these timescales as the time between peak and the luminosity being $1/e$ of the value at peak; we will refer to them as $\tau_r$ and $\tau_d$, respectively. 

The rise and decline timescales are plotted as a function of redshift in Figure~\ref{fig:timescale}, along with the data from \citet{nsj+15b}. Our pseudo-bolometric light curves differ from those of \citet{nsj+15b}, who constructed theirs by summing up the rest-frame ($K$-corrected) $griz$ photometry. This choice would not have been practical for our purposes, since the higher redshifts of our sample mean we lack both sufficiently red spectral coverage, as well as the temporal spectral coverage to calculate $K$-corrections to these filters. In addition, restricting to rest-frame optical would ignore the fact that we do cover the rest-frame UV where the SED peaks, which is one of the unique aspects of our sample. However, this difference means that the timescales derived may also differ somewhat, since the bluest flux also fades the fastest given the temperature evolution (Figure~\ref{fig:bbtemp}). In the two objects overlapping between the samples, PS1-10bzj and PS1-11ap, we recover similar values to within 10\%, however, so this is unlikely to be a significant effect. Typical (statistical) error bars for the rise- and decline timescales are 2-5~days, but as with the peak luminosity this does not capture any systematic effects from our light curves not including the full bolometric light. Generally we find similar time scales in the PS1 sample as in the low-redshift sample, with a few interesting exceptions: PS1-14bj is a clear outlier in both plots, with both the rise and decline being significantly slower than the rest of the sample. PS1-11aib, PS1-11tt and PS1-10ahf show longer rise times than any of the low-redshift objects, though are not nearly as extreme as PS1-14bj; in the case of PS1-11aib the measured rise time is also affected by a possible ``precursor'' bump (Section~\ref{sec:double}). We also note that PS1-14bj and PS1-11aib do not fall on the $\tau_d \simeq 2 \times \tau_r$ correlation found in \citet{nsj+14}, with light curves closer to symmetric in both cases. 

Another interesting feature in Figure~\ref{fig:timescale} is the apparent clustering of decay times into two groups: one fast-declining group with a typical time scale of $30-40$~days, and a slow-declining group with a typical time scale of about 70 days. Whether this is a double-peaked distribution or simply a single-peak distribution with a long tail cannot be determined from the PS1 sample alone, though. We note that PS1-12cil's decline time is intermediate in between the groups, and that PS1-14bj has a significantly longer decline time than any of the other objects in the ``slowly declining'' group, indicating a continuum; this is also supported by other recent compilation studies \citep{ngb17,dgr+17}.

\begin{figure*}
\centering
\begin{tabular}{cc}
\includegraphics{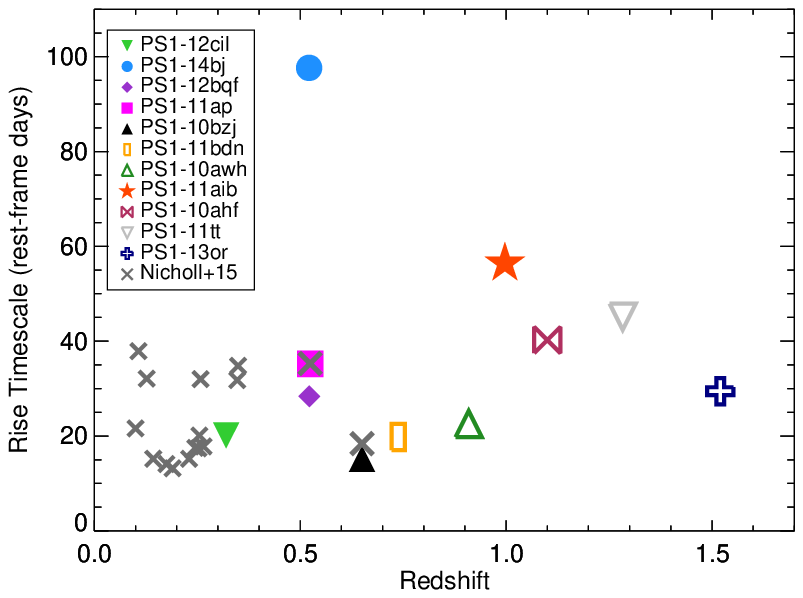} &  \includegraphics{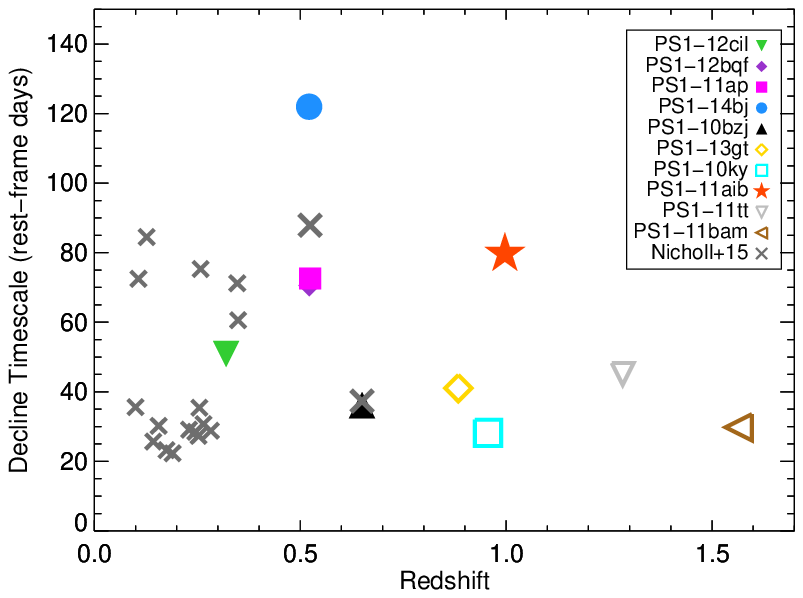}
\end{tabular}
\caption{Rise timescale (left) and decline timescale (right) versus redshift for the PS1 SLSN sample. Gray crosses show the low-redshift sample from \citet{nsj+15b}. PS1-10bzj and PS1-11ap were also analyzed by \citet{nsj+15b}, and we plot the values measured from their \textit{griz}-bolometric light curves as larger crosses for these two objects. Generally the PS1 sample shows similar timescales as the low-redshift objects, with a few exceptions: PS1-14bj is a clear outlier in both plots, showing significantly longer time scales than the rest of the sample. The rise times of PS1-11aib, PS1-10ahf and PS1-11tt are also slower than any of the low-redshift objects.
\label{fig:timescale}}
\end{figure*}

\subsection{Double-peaked Light Curves}
\label{sec:double}
Many SLSN light curves show a double-peaked structure on the rise, with a precursor ``bump'' preceding the main rise. This was first seen in SN\,2006oz \citep{lcd+12} and LSQ14bdq \citep{nsj+15a}, and was suggested to be a ubiquitous feature of H-poor SLSNe by \citet{ns16}. In our sample, there are no clear examples of distinct bumps like that seen in LSQ14bdq, but  PS1-11aib and PS1-13or do show a flattening in their early light curves. The early, marginal \rps detection of PS1-11tt could also be indicative of a precursor (Figure~\ref{fig:lcgold}), but given the sparsely covered rise for this object, the nature of the early detection is unclear. 

Figure~\ref{fig:precursor} shows the rising \ips and \zps light curves of PS1-11aib and PS1-13or. Both objects show structure in the early light curves, in the form of a flattening before rising to the main peak. Compared to the precursor peak seen in LSQ14bdq, these are significantly brighter, with the contrast between the light curve peak and the ``precursor'' less than one magnitude, whereas in LSQ14bdq the contrast was $\sim 2$~mag \citep{nsj+15a}. In fact, precursors like the one in LSQ14bdq would only be detectable in our lowest-redshift data: our typical SLSN peaks at around 22~mag, so a precursor peak as in LSQ14bdq would be $> 24~{\rm mag}$ and thus too faint to be detected. We note that the magnetar shock breakout model of \citet{kmb16} predicted lower contrast between the peaks than was seen in LSQ14bdq; this mechanism might be relevant for PS1-11aib and PS1-13or.

The bottom panels of Figure~\ref{fig:precursor} show the evolution of the blackbody temperature of each event during the rise as calculated in Section~\ref{sec:color}. The temperature evolution is best constrained for PS1-11aib, and shows initial cooling at the beginning of the plateau followed by a flattening out. We note that in the only other SLSN-I with multicolor data available during any kind of precursor event, DES14X3taz, the color and temperature evolution during the precursor was consistent with rapid cooling, and interpreted as shock cooling in extended material \citep{ssd+16}.

PS1-12cil is another object with complex structure in its light curve, but with a second, \textit{late-time} peak or plateau seen in all filters about 40~days after maximum light. The nature of this secondary maximum and its physical interpretation will be discussed in detail in Chornock et al. (2017, in preparation), and we therefore do not investigate it further here.

\begin{figure*}
\centering
\begin{tabular}{cc}
\includegraphics[width=3.5in]{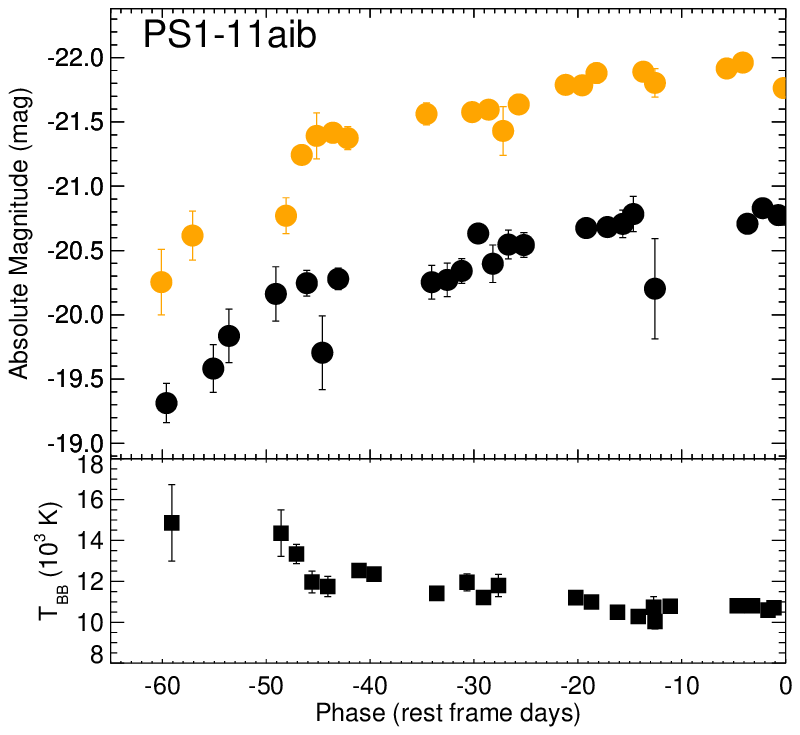} & \includegraphics[width=3.5in]{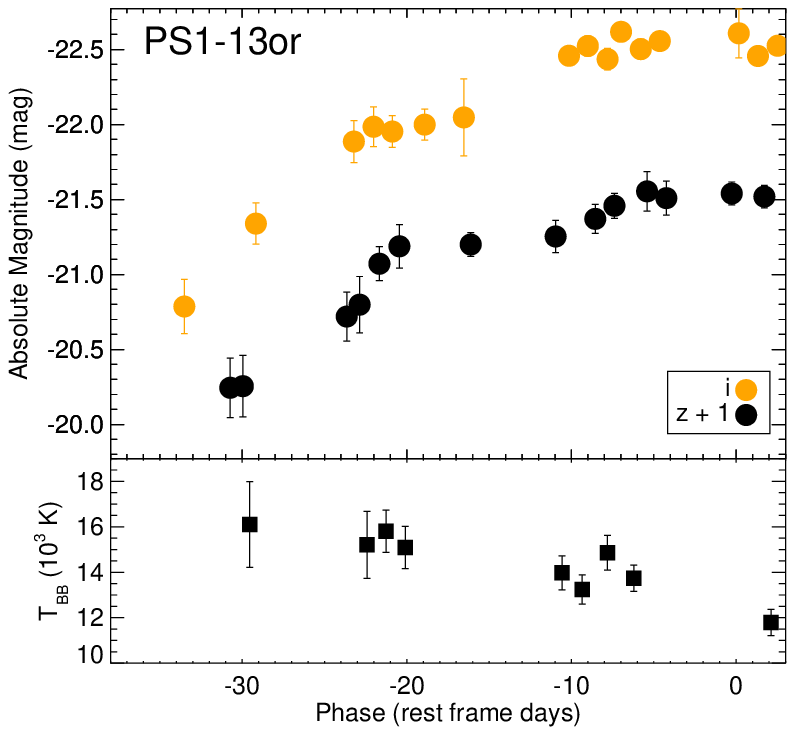}
\end{tabular}
\caption{Zoom-in on the rising \ips and \zps light curves of PS1-11aib (left) and PS1-13or (right). The effective wavelengths of these two filters are approximately 3760~\AA\, and 4340~\AA\, for PS1-11aib and 2980~\AA\, and 3440~\AA\, for PS1-13or (see also Table~\ref{tab:leff}). Both SLSNe show a plateau-like phase in the light curve prior to the main peak. These plateaus are different from the ``prototype'' precursor peak in LSQ14bdq in being significantly brighter and thus with less contrast to the main peak, but demonstrate the diversity and complexity in SLSN light curves. The bottom panels show the evolution of the blackbody temperature.}
\label{fig:precursor}
\end{figure*}

\section{Magnetar Fits to PS1-11aib and PS1-12bqf}
\label{sec:magnetar}
Magnetar spin-down is emerging as a popular model to explain H-poor SLSNe \citep[e.g.,][]{kb10,woo10,dhw+12}, with its flexibility in fitting a variety of light curve shapes, and ability to reproduce the high temperatures and blue spectra seen in SLSNe. Given the varying coverage and quality of our light curves, we do not attempt to model the full set or make inferences about the underlying distributions of parameters assuming a magnetar model. Indeed, the objects from our sample that are previously published already have magnetar fits available in the literature \citep{ccs+11,lcb+13,lcb+16,msk+14,msr+14,ngb17}. Our sample contains two previously unpublished objects, PS1-11aib and PS1-12bqf, with good coverage both on the rise and decline, however, and we explore magnetar model fits to these light curves here. These objects are also interesting in their own rights, with both being slow-decliners, PS1-11aib having a precursor bump, and PS1-12bqf being one of the lowest-luminosity objects in our sample.

Semi-analytic models of magnetar spin-down \citep[e.g.][]{kb10,woo10,isj+13} make the simplifying assumptions of spherical symmetry, magnetic dipole spin-down, and that the energy of the magnetar is thermalized and distributed evenly at the base of the ejecta. The resulting light curve can then be calculated as

\begin{equation}
L_{\rm SN}(t) = \frac{2 E_p}{\tau_p \tau_m}e^{-(\frac{t}{\tau_m})^2}  
\times \int_0^{t} \frac{1}{(1+t'/\tau_p)^2} e^{(\frac{t'}{\tau_m})^2} \frac{t'}{\tau_m} dt' .
\label{eqn:magnetar}
\end{equation}

\noindent Here, $P$ is the initial spin of the magnetar, $B$ is the magnetic field, $E_p \simeq 2 \times 10^{52}~{\rm erg} \times (P/1~{\rm ms})^{-2}$ is the rotational energy of the magnetar, $\tau_p \simeq 4.7~{\rm days} \times (P/1~{\rm ms})^{2} \times (B/10^{14}~{\rm G})^{-2}$ is the spin-down timescale, and $\tau_m$ is the diffusion time. 

A caveat to this approach is that our pseudo-bolometric light curves do not capture the full bolometric light, missing the flux bluewards of our observed bands; the proportion of missed flux will also be higher at early times relative to later times. This caveat will apply to all studies using pseudo-bolometric light curves constructed over a fixed wavelength range; indeed, the fact that our PS1 photometry covers the near-UV in most cases means that we come closer to capturing the full bolometric light than previous studies. A thorough exploration of how different approaches to the bolometric correction would affect the resulting parameters is outside of the scope of this paper, however.

\subsection{PS1-11aib}
As discussed in Section~\ref{sec:double}, the early points on the light curve of PS1-11aib might be part of an early precursor peak similar to what was seen in LSQ14bdq and DES14X3taz \citep{nsj+15b,ssd+16}. Thus, the best fit will depend on whether we attempt to fit the early points as part of the magnetar-powered light curve or not. Figure~\ref{fig:11aib_magnetar} shows several different magnetar fits to the light curve of PS1-11aib. The best three-parameter fit to the full light curve has $M_{\rm ej} \approx 9.8~{\rm M}_{\odot}$, $P \approx 1.9~{\rm ms}$, $B \approx 6 \times 10^{13}~{\rm G}$. This model fits the rise and peak well, but overpredicts the late-time luminosity, which is a common problem with magnetar models. \citet{wwd+15} suggested this could be overcome by accounting for hard emission leakage as the ejecta become transparent to $\gamma$-rays. This is parametrized by adding a term $(1-e^{-A t^{-2}})$ in Equation~\ref{eqn:magnetar}, where $A = 9 \kappa_{\gamma} M_{\rm ej}^2 / 40 \pi E_{\rm K}$ describes the optical depth of the ejecta to gamma rays as $\tau_{\gamma} = A t^{-2}$.  Larger values of $A$ correspond to a larger trapping rate and lower leakage rate; the original magnetar model has  $A = \infty$. When adding a hard emission leakage term of $A = 2 \times 10^{14}~{\rm s}^2$, our best-fit model can also account for the late-time data point.

If we interpret the early light curve as a precursor peak, and powered by a different mechanism than the main light curve, the effective rise time of the magnetar-powered light curve is shorter. The best fit when excluding the early data ($>35$~days prior to peak) has a higher magnetic field ($B = 10^{14}~{\rm G}$) and a slightly lower ejecta mass ($M_{\rm ej} = 8~{\rm M}_{\odot}$) compared to the model above, both of which contribute to making the light curve narrower; the initial spin is similar ($P = 2.0~{\rm ms}$). In this case, it is not necessary to invoke late-time leakage to fit the data point at $+100$~days.

\begin{figure}
\centering
\includegraphics[width=3.5in]{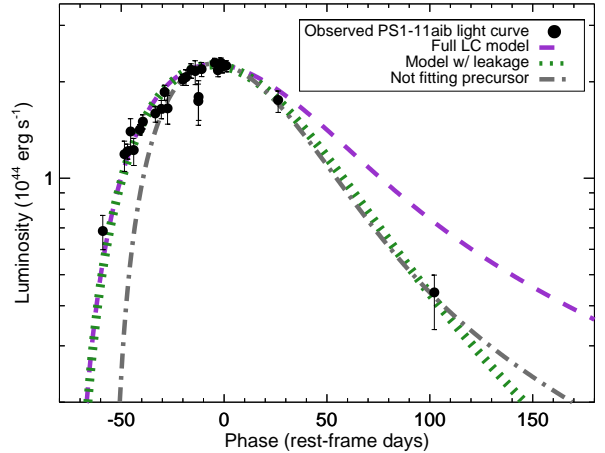}
\caption{Example magnetar model fits to the bolometric light curve of PS1-11aib. The rise and peak are reasonably well fit by a simple magnetar model (purple dashed curve), but the late-time luminosity is overpredicted; this can be mitigated by including late-time hard emission leakage (green dotted curve). Alternatively, if the early emission is interpreted as part of a precursor peak, the peak and late-time data are well fit by a simple magnetar model with a shorter rise time (gray dot-dashed curve). 
\label{fig:11aib_magnetar}}
\end{figure}

\subsection{PS1-12bqf}

Figure~\ref{fig:12bqf_magnetar} shows the best-fit magnetar model to the light curve of PS1-12bqf. Unlike PS1-11aib, we do not need to invoke hard emission leakage, and the light curve is well fit with a simple model with $M_{\rm ej} = 3~M_{\odot}$, $P = 4.8~{\rm ms}$, and $B= 1 \times 10^{14}~{\rm G}$. Given the noise in the light curve, we can find adequate fits with a range of parameters; faster initial spin periods also require higher magnetic fields to reproduce the peak luminosity and rise time. Although PS1-12bqf is lower-luminosity than most H-poor SLSNe, the fact that it can be well fit with a magnetar should not be surprising: magnetar models have a large parameter space, and can naturally produce light curves with a range of luminosities. We note that the values we derive for PS1-12bqf are within the distribution of parameters found by \citet{ngb17}, and similar to what they derive for the SLSNe PTF10hgi and LSQ14mo. Our code is not set up to do a full parameter exploration and calculate confidence intervals; we refer the reader to recent Markov-Chain Monte Carlo (MCMC) efforts to model H-poor SLSNe for typical parameter ranges \citep{ngb17,lww+17,gnv+17}. Such MCMC efforts are also better suited to explore degeneracies and covariances between the different parameters.

\begin{figure}
\centering
\includegraphics[width=3.5in]{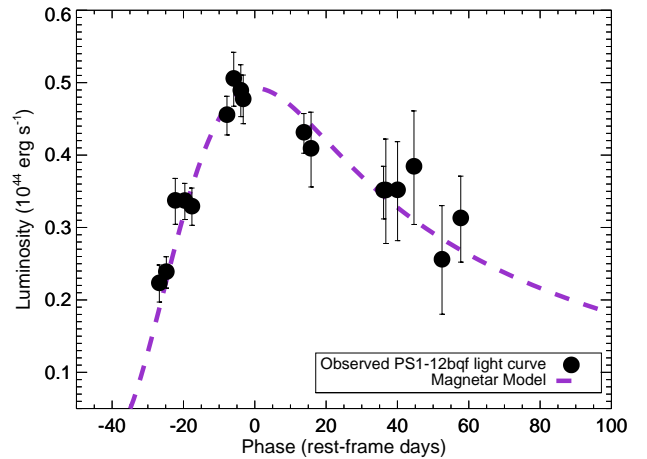}
\caption{Magnetar model fit to the bolometric light curve of PS1-12bqf. Although lower-luminosity than most objects in our sample, the magnetar spin-down model also easily reproduces light curves in this part of parameter space.}
\label{fig:12bqf_magnetar}
\end{figure}

\section{Luminosity Function of H-poor SLSNe}
\label{sec:comp_disc}

It has been claimed that H-poor SLSNe show a tight distribution in peak luminosities, as well as correlations between peak luminosity, color, and decline rates \citep{is14}, and as a result may be useful standardizable candles for cosmology. SLSNe as standard candles is an attractive idea, given that the high UV luminosities make them detectable to much higher redshifts than Type Ia SNe. While the initial study by \citet{is14} utilized a heterogeneous data set with objects from many different surveys, the PS1~MDS sample has the advantage of a uniform survey with a well-defined footprint and cadence. While the PS1~MDS supernova sample is far from spectroscopically complete, we can investigate the relative distributions of peak luminosities, with the underlying simplifying assumption that the likelihood that a particular SLSN is observed spectroscopically depends only on its apparent magnitude (Malmquist bias), but not on any other intrinsic properties we wish to compare.

In order to compare the peak luminosities across the wide redshift range of the PS1~MDS sample, we need to $K$-correct them to a common bandpass. For this purpose, we choose the same fiducial bandpass centered at 4000~\AA\, as was used in \citet{is14}, both for the purpose of comparing the scatter directly, and because the claim was that this was a spectral region relatively free of strong spectral features and thus well-suited for such comparisons. In addition, $\sim 4000$~\AA\, is roughly the longest wavelength that is covered by our photometric data over the entire redshift range. For each SLSN, we pick the filter closest to the rest wavelength of 4000~\AA\,, and calculate the $K$-correction from this filter into the fiducial bandpass using either the spectrum near peak (where possible) or a blackbody function constructed from the photometry at peak. We use {\tt SNAKE}\footnote{\url https://github.com/cinserra/S3} \citep{isg+16} to calculate the $K$-correction. 

In Figure~\ref{fig:absmag_vs_z} we show the resulting peak absolute magnitudes in the 4000~\AA\, fiducial bandpass, plotted as a function of redshift. To illustrate the effects of a flux-limited survey, we also plot the absolute magnitude as a function of redshift corresponding to an apparent magnitude of 23.5~mag, the limiting magnitude of PS1~MDS nightly images. In practice, the more relevant flux limit comes from the requirement of spectroscopic classification: given that our spectroscopic follow-up resources were 6- and 8-m class telescopes, we rarely took spectra of objects fainter than 22.5~mag. A striking feature of Figure~\ref{fig:absmag_vs_z} is the spread in luminosities, showing that the SLSN luminosity function clearly extends from at least $-20.5$~mag to about $-22.5$~mag. Again, this indicates that a strict luminosity cutoff is not a suitable way to select SLSNe. We also note that the peak luminosities of our faintest SLSNe are comparable to some of most luminous Type Ic-BL SNe discovered (e.g., \citealt{wlk+17,ssv+12}), suggesting that there is not a true luminosity ``gap'' between SLSNe and core-collapse SNe. 

We plot the cumulative distribution of absolute magnitudes in Figure~\ref{fig:lumdist}. PS1-13gt is included in the distribution as a lower limit, since we only observe its decline and not the peak. We show the Kaplan-Meier estimator of observed peak absolute magnitudes, as well as the volume-corrected distribution, where each observation is weighted by $1/V_{\rm max}$, where $V_{\rm max}$ is the maximum volume each SLSN could have been observed to given its absolute magnitude and an assumed flux limit of $22.5~{\rm mag}$. We also correct for that the effective survey length is different at different redshifts due to time dilation.
Taking these effects into account shifts the median absolute magnitude to $-21.1~{\rm mag}$, from $-21.8~{\rm mag}$ in the uncorrected distribution. This is significantly fainter than the mean magnitude of $-21.7~{\rm mag}$ found in the study of \citet{is14} in the same bandpass, which is unsurprising since their study was unable to account for Malmquist bias. Thus, the results from the PS1~MDS sample suggest that lower-luminosity SLSNe are intrinsically common. 

Another interesting question is the scatter in the luminosity distribution, a key quantity for assessing the utility of SLSNe as standard candles. In the case of the PS1~MDS sample, we find a mean and standard deviation of $-21.70 \pm 0.72$~mag in the observed sample, and $-21.31 \pm 0.73$ in the volume-weighted sample, respectively; both estimates exclude PS1-13gt. \citet{is14}, by comparison, found a raw scatter of $0.46~{\rm mag}$ in peak $M_{4000}$ magnitudes in the similarly-sized literature sample they considered. The PS1~MDS sample therefore does not reproduce the initial findings that SLSNe show a low intrinsic scatter. We note that our results both in terms of mean luminosity and scatter are very similar to that found in the independent, lower-redshift PTF sample ($\langle M_g \rangle = -21.14~{\rm mag}$; scatter $\sigma = 0.74~{\rm mag}$; \citealt{dgr+17}), supporting our findings and also suggesting the SLSN luminosity function does not evolve significantly over this redshift range.

\citet{is14} also showed that the scatter in their sample was further reduced by considering correlations with decline time and color. Unfortunately, we do not have the wavelength coverage to perform the color comparisons, or the spectral coverage to calculate accurate late-time $K$-corrections to measure the decline rates. We note, however, that several objects in the PS1~MDS sample do not follow the trend they find that higher-luminosity objects have broader light curves: two of our faintest objects, PS1-12bqf and PS1-14bj are both slow decliners, while several of the brightest objects in the sample, like PS1-11bam, have fast decline timescales. 

In addition to the peak $M_{4000}$ magnitudes, we use the spectra to also calculate peak absolute magnitudes at a rest-frame wavelength of 2600~\AA\, (again using SNAKE, $K$-correcting to the rest-frame {\it Swift} $uvw1$ bandpass). If we do not have a sufficiently blue spectrum available for the object in question, we calculate the $K$-correction using the spectrum of an object of similar temperature: the spectrum of PS1-10bzj was used for PS1-12cil, and the spectrum of PS1-11aib for PS1-10ahf and PS1-10pm. PS1-13gt is not included in this plot, as we do not have an appropriate spectrum available. The resulting luminosity distribution is shown as a cumulative histogram in Figure~\ref{fig:lumdist_2600}. The general trend is similar to at 4000~\AA\,; we find a median peak magnitude of $-21.1~{\rm mag}$ in the volume-corrected sample compared to $-21.9~{\rm mag}$ in the uncorrected sample. Both the overall range and the spread are larger in the UV, however: we find a mean and standard deviation of $-20.9 \pm 1.15 ~{\rm mag}$ in the volume-weighted sample. This larger spread reflects the actual temperature variations (Figure~\ref{fig:bbtemp}), the strength of the absorption features present in this wavelength region, and possibly also variations in dust extinction. We note the overall high UV luminosities, which make SLSNe excellent targets for UV spectroscopy, allowing for studying both the SLSNe themselves and their host galaxy environments through absorption spectroscopy (e.g., \citealt{qkk+11,bcl+12,vsg+14,yqg+17}).

Finally, we note that none of the objects found in PS1~MDS are more luminous than $-23~{\rm mag}$, despite the large volume over which we would be sensitive to such objects. This suggests that there is an upper cutoff to the SLSN luminosity function, or at least that such luminous SLSNe have to be intrinsically rare. If an object like ASASSN-15lh, which peaked at $-23.5~{\rm mag}$, were indeed a SLSN \citep{dsp+16,gsk+17}, it would be significantly more extreme than any of the objects found in PS1~MDS. Given this transient's location at the center of a massive galaxy (while SLSN-I are almost exclusively found in low-mass, low-metallicity galaxies; e.g., \citealt{lcb+14,lcb+15,lsk+15,pqy+16}), it has also been suggested that this transient was a tidal disruption event rather than a true SLSN \citep{pqy+16,lfs+17,mmc+17}.

\begin{figure}
\centering
\includegraphics{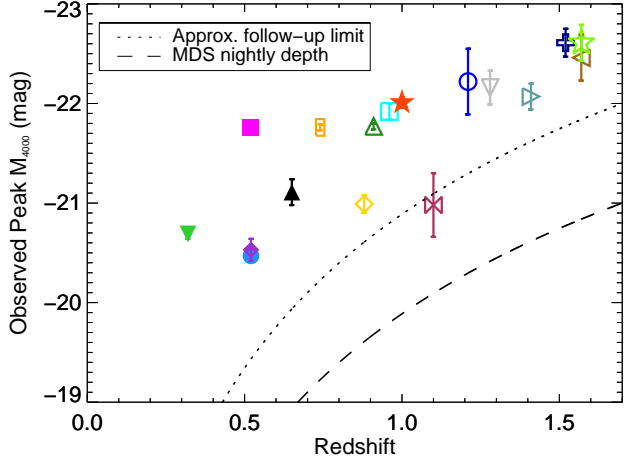}
\caption{$M_{4000}$, absolute magnitude in the 4000~\AA\, bandpass used by \citet{is14} versus redshift for the PS1~MDS sample. Plot symbols for individual SLSNe are the same as in Figures~\ref{fig:bbtemp}-\ref{fig:timescale}. The dashed line shows the limiting magnitude of PS1~MDS nightly images, while the dotted line is our effective survey depth for the spectroscopic sample.}
\label{fig:absmag_vs_z}
\end{figure}

\begin{figure}
\centering
\includegraphics[width=3.5in]{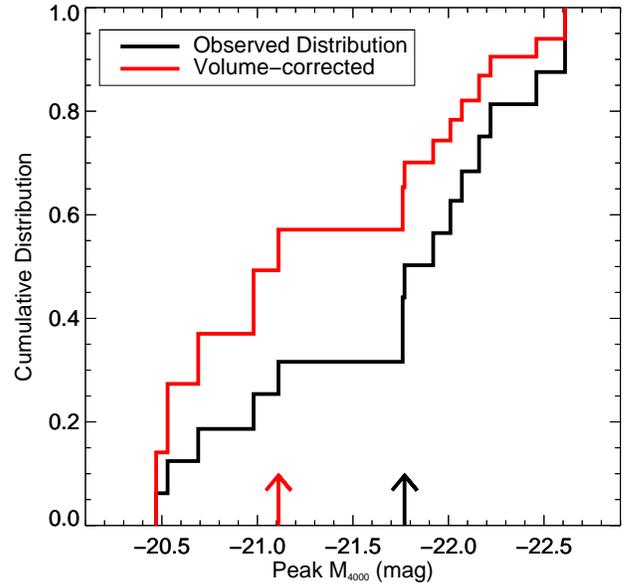}
\caption{Cumulative distribution of peak absolute magnitudes in the 4000~\AA\, bandpass. The black line shows the distribution as observed, with a median value of $-21.9~{\rm mag}$ (shown by the black arrow). The red line shows the resulting distribution when accounting for effective volume and survey time for each SLSN, bringing the median to $-21.1~{\rm mag}$ (shown by the red arrow). 
\label{fig:lumdist}
}
\end{figure}

\begin{figure}
\centering
\includegraphics[width=3.5in]{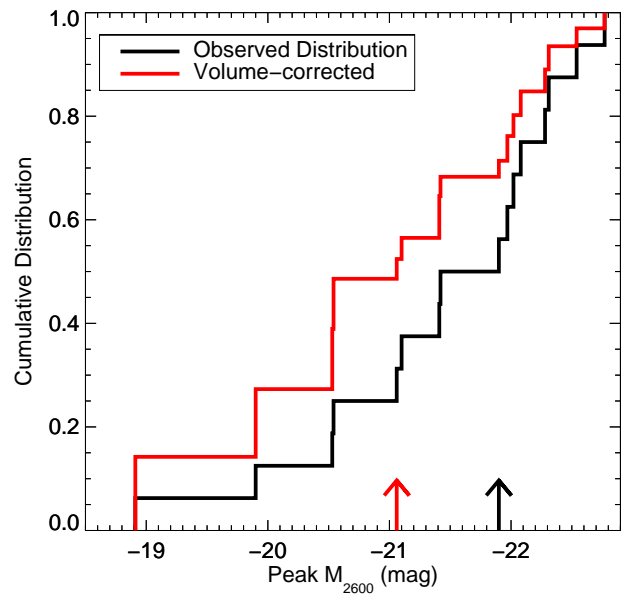}
\caption{Like Figure~\ref{fig:lumdist}, but with absolute magnitudes calculated at an effective wavelength of 2600~\AA\, ({\it Swift} UVOT $uvw1$ band). The trends are similar to at 4000~\AA\,, but the observed range and spread in absolute magnitudes are larger at this wavelength.
\label{fig:lumdist_2600}
}
\end{figure}

\section{Conclusions}
\label{sec:comp_conc}
We have presented the light curves and classification spectra of 17 H-poor SLSNe from PS1~MDS. Our sample contains all objects that are spectroscopically similar to either of the prototypical objects SN\,2005ap/SCP06F6 or SN\,2007bi, without an explicit limit on luminosity. With a median redshift of $z \sim 1$, this is the largest sample of high-redshift SLSNe presented to date. Utilizing the light curves and spectra, our findings can be summarized as follows:

\begin{itemize}
\item The light curves of H-poor SLSNe are diverse. The lower limits on peak bolometric luminosities in our sample span $(0.5 - 5) \times 10^{44}~{\rm erg~s}^{-1}$, measured rise timescales $15-95~{\rm days}$, and decline timescales $30-135~{\rm days}$. Similarly, the lower limits on total radiated energy for our sample span $(0.3 - 2) \times 10^{51}~{\rm erg}$. 
\item Prior to peak light, H-poor SLSNe show hot color temperatures ($10,000 - 25,000$~K) over a timescale of weeks, suggesting there is a sustained source of heating. Post-peak, color temperatures drop to $\sim 6000-8000~{\rm K}$ over a timescale of $20-40~{\rm days}$. 
\item At least two objects (PS1-11aib and PS1-13or) show plateaus in the early light curves, both of which are significantly brighter than the precursor peaks that have been seen in typical low-redshift objects like LSQ14bdq. The temperature evolution of PS1-11aib shows initial cooling at the beginning of the plateau, then a flat temperature.
\item Our spectroscopically selected sample contains several objects with peak luminosities fainter than $-21~{\rm mag}$, suggesting that such a luminosity cut is arbitrary. After correcting for the effective volume probed by each SLSN, we find a median peak magnitude at 4000~\AA\, of $-21.1$~mag. We find an intrinsic spread in peak magnitudes of $\sim 0.7$~mag, higher than previous studies compiled from literature data. At 2600~\AA\,, we find a median peak magnitude of $-21.1~{\rm mag}$ and a larger scatter of $1.2~{\rm mag}$.
\end{itemize}

Our results highlight the need for a better understanding of sample selection when discussing the properties of SLSNe as a class, both in terms of survey biases and in terms of which objects are reported as superluminous. The luminosity function derived from the spectroscopically selected PS1~MDS sample shows that a luminosity threshold is not an appropriate way to select SLSNe, and may exclude a large fraction of the true population. Similarly, the large scatter in luminosities and diversity in light curve shapes indicate that the utility of H-poor SLSNe for cosmology may be limited. While this diversity complicates the selection of SLSNe from large upcoming surveys like LSST, the PS1~MDS data set will serve as a valuable training set.

\acknowledgements
We thank the anonymous referee for constructive comments that improved the manuscript. R.L. thanks Robert Quimby for helpful comments on an early version of this draft, and acknowledges helpful discussions at the MIAPP workshop ``Superluminous Supernovae in the Next Decade'', supported by the Munich Institute for Astro- and Particle Physics (MIAPP) of the DFG cluster of excellence ``Origin and Structure of the Universe''.
The Pan-STARRS1 Surveys (PS1) have been made possible through contributions by the Institute for Astronomy, the University of Hawaii, the Pan-STARRS Project Office, the Max-Planck Society and its participating institutes, the Max Planck Institute for Astronomy, Heidelberg and the Max Planck Institute for Extraterrestrial Physics, Garching, The Johns Hopkins University, Durham University, the University of Edinburgh, the Queen's University Belfast, the Harvard-Smithsonian Center for Astrophysics, the Las Cumbres Observatory Global Telescope Network Incorporated, the National Central University of Taiwan, the Space Telescope Science Institute, and the National Aeronautics and Space Administration under Grant No. NNX08AR22G issued through the Planetary Science Division of the NASA Science Mission Directorate, the National Science Foundation Grant No. AST-1238877, the University of Maryland, Eotvos Lorand University (ELTE), and the Los Alamos National Laboratory. 
This work is based on observations obtained at the Gemini Observatory, which is operated by the Association of Universities for Research in Astronomy, Inc., under a cooperative agreement with the NSF on behalf of the Gemini partnership: the National Science Foundation (United States), the National Research Council (Canada), CONICYT (Chile), the Australian Research Council (Australia), Minist\'{e}rio da Ci\^{e}ncia, Tecnologia e Inova\c{c}\~{a}o (Brazil) and Ministerio de Ciencia, Tecnolog\'{i}a e Innovaci\'{o}n Productiva (Argentina). 
This work includes data gathered with the 6.5 meter Magellan Telescopes located at Las Campanas Observatory, Chile. Observations reported here were obtained at the MMT Observatory, a joint facility of the Smithsonian Institution and the University of Arizona. 
Some of the computations in this paper were run on the Odyssey cluster supported by the FAS Division of Science, Research Computing Group at Harvard University. The UCSC group is supported in part by NSF grant AST-1518052, the Gordon \& Betty Moore Foundation, and from fellowships from the Alfred P.\ Sloan Foundation and the David and Lucile Packard Foundation to R.J.F. SJS. acknowledges funding from the European Research Council under the European Union's Seventh Framework Programme (FP7/2007-2013)/ERC Grant agreement n$^{\rm o}$ [291222]
and STFC grants ST/I001123/1 and ST/L000709/1. 
R.C. thanks the Kavli Institute for Theoretical Physics for its hospitality while this work was in the final stages of preparation. This research was supported in part by the National Science Foundation under grant no. NSF PHY11-25915.
T.L. is a Jansky Fellow of the National Radio Astronomy Observatory.
This research has made use of NASA's Astrophysics Data System.

\textit{Facilities:} \facility{PS1}, \facility{MMT}, \facility{Magellan:Clay}, \facility{Magellan:Baade}, \facility{Gemini:Gillett}, \facility{Gemini:South}

\clearpage
\newpage
\begin{deluxetable*}{lcccc}
\tablewidth{0pt}
\tabletypesize{\scriptsize}
\tablecaption{SLSNe from PS1~MDS}
\tablehead{
\colhead{Object} &
\colhead{Redshift} &
\colhead{RA} &
\colhead{Dec} &
\colhead{Reference}
}
\startdata
PS1-12cil   & 0.32   & \ra{08}{40}{56.169} & \dec{+45}{24}{41.93}  & \\
PS1-14bj    & 0.5125    & \ra{10}{02}{08.433} & \dec{+03}{39}{19.02}    &  \citet{lcb+16} \\
PS1-12bqf   & 0.522   & \ra{02}{24}{54.621} & \dec{-04}{50}{22.72}    &   \\
PS1-11ap    & 0.524   & \ra{10}{48}{27.752} & \dec{+57}{09}{09.32}    & \citet{msk+14}  \\
PS1-10bzj   & 0.650   & \ra{03}{31}{39.826} & \dec{-27}{47}{42.17}    & \citet{lcb+13}  \\
PS1-11bdn   & 0.738   & \ra{02}{25}{46.292} & \dec{-05}{03}{56.57}    &   \\
PS1-13gt    & 0.884   & \ra{12}{18}{02.035} & \dec{+47}{34}{45.95}    &   \\
PS1-10awh   & 0.909   & \ra{22}{14}{29.831} & \dec{-00}{04}{03.62}    & \citet{ccs+11}  \\
PS1-10ky    & 0.956   & \ra{22}{13}{37.851} & \dec{+01}{14}{23.57}    & \citet{ccs+11}  \\
PS1-11aib   & 0.997   & \ra{22}{18}{12.217} & \dec{+01}{33}{32.01}    &   \\
PS1-10ahf   & 1.10   & \ra{23}{32}{28.311} & \dec{-00}{21}{43.46}    & \citet{msr+14}  \\
PS1-10pm    & 1.206   & \ra{12}{12}{42.200} & \dec{+46}{59}{29.48}    & \citet{msr+14}  \\
PS1-11tt    & 1.283   & \ra{16}{12}{45.778} & \dec{+54}{04}{16.96}    &   \\
PS1-11afv   & 1.407   & \ra{12}{15}{37.770} & \dec{+48}{10}{48.62}    &   \\
PS1-13or    & 1.52    & \ra{09}{54}{40.296} & \dec{+02}{11}{42.24}    &   \\
PS1-11bam   & 1.565   & \ra{08}{41}{14.192} & \dec{+44}{01}{56.95}    & \citet{bcl+12}  \\
PS1-12bmy   & 1.572   & \ra{03}{34}{13.123} & \dec{-26}{31}{17.21}    &
\enddata
\label{tab:targets}
\end{deluxetable*}

\begin{deluxetable*}{lccccc}
\tablewidth{0pt}
\tabletypesize{\scriptsize}
\tablecaption{Effective Wavelengths of PS1 Bandpasses}
\tablehead{
\colhead{Object} &
\colhead{\gps} &
\colhead{\rps} &
\colhead{\ips} &
\colhead{\zps} &
\colhead{\yps} \\
\colhead{} &
\colhead{(\AA)} &
\colhead{(\AA)} &
\colhead{(\AA)} &
\colhead{(\AA)} &
\colhead{(\AA)} 
}
\startdata
PS1-12cil & 3647 & 4677 & 5693 & 6563 & 7285 \\
PS1-12bqf & 3163 & 4056 & 4938 & 5692 & 6318 \\
 PS1-11ap & 3158 & 4051 & 4931 & 5684 & 6310 \\
 PS1-14bj & 3164 & 4058 & 4939 & 5694 & 6320 \\
PS1-10bzj & 2917 & 3742 & 4555 & 5250 & 5828 \\
PS1-11bdn & 2770 & 3552 & 4324 & 4984 & 5533 \\
 PS1-13gt & 2555 & 3277 & 3989 & 4598 & 5104 \\
PS1-10awh & 2521 & 3234 & 3937 & 4538 & 5037 \\
 PS1-10ky & 2461 & 3156 & 3842 & 4429 & 4916 \\
PS1-11aib & 2410 & 3091 & 3763 & 4338 & 4815 \\
PS1-10ahf & 2230 & 2861 & 3482 & 4014 & 4456 \\
 PS1-10pm & 2182 & 2798 & 3406 & 3927 & 4359 \\
 PS1-11tt & 2108 & 2704 & 3292 & 3794 & 4212 \\
PS1-11afv & 2000 & 2565 & 3122 & 3599 & 3995 \\
 PS1-13or & 1910 & 2450 & 2982 & 3437 & 3816 \\
PS1-11bam & 1876 & 2407 & 2930 & 3377 & 3749 \\
PS1-12bmy & 1871 & 2400 & 2922 & 3368 & 3739 
\enddata
\label{tab:leff}
\end{deluxetable*}

\begin{deluxetable*}{lcccccccccc}
\tablewidth{0pt}
\tabletypesize{\scriptsize}
\tablecaption{Log of Spectroscopic Observations}
\tablehead{
\colhead{Object} &
\colhead{UT Date} &
\colhead{Phase} &
\colhead{Instrument} &
\colhead{Wavelength Range} &
\colhead{Slit} &
\colhead{Grating} &
\colhead{Filter} &
\colhead{Exp. time} &
\colhead{Airmass} \\
\colhead{} &
\colhead{(YYYY-MM-DD.D)} &
\colhead{(days)} &
\colhead{} &
\colhead{(\AA)} &
\colhead{(\arcsec)} &
\colhead{} &
\colhead{} &
\colhead{(s)} &
\colhead{}
}
\startdata
PS1-12cil  &  2013-01-12.3 &   $+1$  & MMT/BlueChannel & 3310-8520  & 2.0 & 300GPM & none & 2700 & 1.1  \\
PS1-12bqf  &  2012-11-14.2 &   $-1$  & MMT/BlueChannel & 3310-8530  & 1.0 & 300GPM & none & 3000 & 1.2  \\
PS1-11bdn  &  2012-01-01.1 &   $-8$  & MMT/BlueChannel & 3370-8580  & 1.0  & 300GPM & none & 3600 & 1.3 \\
PS1-13gt   &  2013-03-05.7 & \nodata & GN/GMOS         & 5880-10160 & 1.0  & R400  & OG515 & 3600  & 1.1\\
PS1-11aib  &  2011-11-28.1 &   $+16$ & MMT/BlueChannel & 3330-8540  & 1.0  & 300GPM & none & 5400 & 1.3  \\
PS1-11tt   &  2011-06-07.5 & $+4$    & GN/GMOS         & 4860-8640  & 1.0  & R400   & GG455 & 3000 & 1.5 \\
PS1-11afv  &  2011-07-09.3 & $+9$     & GN/GMOS         & 4900-9150  & 1.0  & R400   & GG455 & 2400 & 1.5 \\
PS1-13or   &  2013-05-04.0 & $+2$ & GS/GMOS         & 4890-9140  & 1.0  & R400   & GG455 & 3600 & 1.2 \\
PS1-12bmy  &  2012-11-11.1 & $+5$ & GS/GMOS         & 4890-9140  & 1.0  & R400   & GG455 & 3600 & 1.1
\enddata
\label{tab:spectra}
\end{deluxetable*}

\begin{deluxetable*}{lccccc}
\tablewidth{0pt}
\tabletypesize{\scriptsize}
\tablecaption{Photometry of PS1 SLSNe}
\tablehead{
\colhead{Object} &
\colhead{MJD} & 
\colhead{Rest-frame Phase} & 
\colhead{Filter} &
\colhead{AB Mag} &
\colhead{Instrument} \\
\colhead{} &
\colhead{(days)} &
\colhead{(days)}  & 
\colhead{} &
\colhead{} &
\colhead{} 
}
\startdata
PS1-12bqf  &  56206.6  &  $-26.8$  &  g$_{\rm P1}$  &  22.48 $\pm$ 0.14  &  PS1   \\
PS1-12bqf  &  56209.6  &  $-24.8$  &  g$_{\rm P1}$  &  22.22 $\pm$ 0.09  &  PS1   \\
PS1-12bqf  &  56214.4  &  $-21.7$  &  g$_{\rm P1}$  &  22.09 $\pm$ 0.11  &  PS1   \\
PS1-12bqf  &  56217.5  &  $-19.6$  &  g$_{\rm P1}$  &  21.85 $\pm$ 0.07  &  PS1   \\
PS1-12bqf  &  56220.5  &  $-17.6$  &  g$_{\rm P1}$  &  22.11 $\pm$ 0.12  &  PS1   \\
PS1-12bqf  &  56235.5  &  $-7.8$  &  g$_{\rm P1}$  &  21.76 $\pm$ 0.08  &  PS1   \\
PS1-12bqf  &  56238.3  &  $-5.9$  &  g$_{\rm P1}$  &  21.56 $\pm$ 0.11  &  PS1   \\
PS1-12bqf  &  56241.4  &  $-3.9$  &  g$_{\rm P1}$  &  21.73 $\pm$ 0.12  &  PS1   \\
PS1-12bqf  &  56268.3  &  $13.8$  &  g$_{\rm P1}$  &  22.38 $\pm$ 0.12  &  PS1   \\
PS1-12bqf  &  56271.3  &  $15.8$  &  g$_{\rm P1}$  &  22.17 $\pm$ 0.08  &  PS1   
\enddata
\label{tab:phot}
\tablecomments{The full table is included as a separate file in this posting.}
\end{deluxetable*}

\begin{deluxetable*}{lcccccccc}
\tablewidth{0pt}
\tabletypesize{\scriptsize}
\tablecaption{Derived Properties}
\tablehead{
\colhead{Object} &
\colhead{T$_{\rm BB}$ at peak} &
\colhead{Peak Lum.} & 
\colhead{Rad. Energy\tablenotemark{a}} & 
\colhead{$\tau_r$} &
\colhead{$\tau_d$} &
\colhead{Velocity at peak\tablenotemark{b}} &
\colhead{M$_{400}$} &
\colhead{M$_{260}$} \\
\colhead{} &
\colhead{(K)} &
\colhead{($10^{44}~{\rm erg/s}$)} &
\colhead{($10^{51}~{\rm erg}$)}  & 
\colhead{(days)} &
\colhead{(days)} &
\colhead{(km~s$^{-1}$)} &
\colhead{(AB mag)} &
\colhead{(AB mag)}
}
\startdata
PS1-12cil  &  13,000  &   0.50  &   0.22  &  20.1  &   50.9  &  \nodata  &  $-20.69 \pm 0.05$ & $-20.54 \pm 0.05$ \\
 PS1-14bj  &  7,000  &   0.46  &   0.78  &  97.6  &  122.0  &  5,000\tablenotemark{c}  &  $-20.47 \pm 0.04$ & $-18.91 \pm 0.06$ \\
PS1-12bqf  &  11,000  &   0.47  &   0.28  &  28.4  &   70.6  &  14,000 & $-20.53 \pm 0.11$ &  $-19.90 \pm 0.14$ \\
 PS1-11ap  &  10,000  &   1.63  &   1.04  &  35.2  &   72.6  &  16,000  & $-21.86 \pm 0.05$ & $-21.06 \pm 0.10$ \\
PS1-10bzj  &  17,000  &   1.17  &   0.37  &  15.2  &   36.1  &  14,000  & $-21.11 \pm 0.13$ &  $-21.41 \pm 0.17$ \\
PS1-11bdn  &  12,000  &   4.70  &   0.61  &  19.8  &  \nodata  &  16,000 & $-21.76 \pm 0.03$ &  $-22.31 \pm 0.07$ \\
 PS1-13gt  &  6,000  &   1.25  &   0.40  &  \nodata  &   41.0  &  \nodata & $-20.99 \pm 0.09$ & \nodata  \\
PS1-10awh  &  16,000  &   2.16  &   0.59  &  22.5  &  \nodata  &  13,000  &  $-21.77 \pm 0.03$ &  $-21.97 \pm 0.10$ \\
 PS1-10ky  &  16,000  &   2.75  &   0.58  &  \nodata  &   28.2  &  18,000 &  $-21.92 \pm 0.08$ & $-22.28 \pm 0.13$ \\
PS1-11aib  &  10,000  &   2.24  &   2.02  &  56.5  &   79.8  &  16,000 & $-22.01 \pm 0.05$ &  $-22.02 \pm 0.17$ \\
 PS1-10pm  &  8,000  &   2.56  &   0.77  &  \nodata  &  \nodata  &  16,000 & $-22.22 \pm 0.33$ & $-21.42 \pm 0.10$ \\
 PS1-11tt  &  9,000  &   2.61  &   1.19  &  45.2  &   45.1  &   9,000  &  $-22.16 \pm 0.17$ & $-21.10 \pm 0.15$ \\
PS1-11afv  &  12,000  &   2.32  &   0.41  &  \nodata  &  \nodata  &   9,000  &  $-22.07 \pm 0.13$ &  $-22.08 \pm 0.16$ \\
 PS1-13or  &  11,000  &   5.20  &   1.12  &  29.5  &  \nodata  &  \nodata  &  $-22.61 \pm 0.14$ &  $-22.77 \pm 0.20$ \\
PS1-11bam  &  12,000  &   4.13  &   0.94  &  \nodata  &   29.8  &  17,000  &  $-22.46 \pm 0.23$ &  $-22.54 \pm 0.10$ \\
PS1-12bmy  &  9,000  &   3.51  &   1.04  &  \nodata  &   30.5  &  16,000   & $-22.61 \pm 0.18$ & $-21.90 \pm 0.18$ 
\enddata
\tablenotetext{a}{Lower limits.}
\tablenotetext{b}{Measured from the minimum of the \ion{Mg}{2} feature, unless stated otherwise.}
\tablenotetext{c}{From the SYNOW fit presented in \citet{lcb+16}.}
\label{tab:results}
\end{deluxetable*}

\end{document}